\documentclass[fleqn,usenatbib]{mnras}

\usepackage{newtxtext,newtxmath}

\usepackage[T1]{fontenc}
\usepackage{orcidlink}
\DeclareRobustCommand{\VAN}[3]{#2}
\let\VANthebibliography\thebibliography
\def\thebibliography{\DeclareRobustCommand{\VAN}[3]{##3}\VANthebibliography}

\usepackage[export]{adjustbox}

\usepackage{graphicx}
\usepackage{amsmath}	
\usepackage{natbib}
\defcitealias{presa24}{P24}
\usepackage{enumitem}
\usepackage{float}

\title[Magnetic coupling and evaporating atmospheres]{Star-planet magnetic interactions in photoevaporating exoplanets: enhanced power due to atmospheric escape}

\author[A. Presa, A. A. Vidotto and F. Elekes]{
Andrés Presa \orcidlink{0009-0006-0590-9346},\thanks{E-mail: presa@strw.leidenuniv.nl}
Aline A. Vidotto \orcidlink{0000-0001-5371-2675},
Filip Elekes \orcidlink{0000-0002-7258-3386}
\\
Leiden Observatory, Leiden University, PO Box 9513, 2300 RA, Leiden, The Netherlands\\
}

\date{Accepted XXX. Received YYY; in original form ZZZ}

\pubyear{\the\year{}}

\begin{document}
\label{firstpage}
\pagerange{\pageref{firstpage}--\pageref{lastpage}}
\maketitle

\begin{abstract}
Observations of periodic stellar activity near the transit phase of a close-in exoplanet provide evidence of star-planet magnetic interactions (SPMI), similar to the magnetic coupling between Jupiter and its moons. Comparing the power associated with SPMI signals to analytical theories offers a way to constrain exoplanetary magnetic fields, but models based on moon–magnetosphere analogs often underpredict observed energy fluxes. Unlike moons, many close-in exoplanets are extended, highly irradiated gas giants undergoing significant photoevaporation. However, it is not known how atmospheric escape influences the star-planet magnetic coupling. Here, we present three-dimensional radiation magneto-hydrodynamic simulations that simultaneously model planetary evaporation and SPMI in a hot Jupiter planet embedded in a magnetised stellar wind. Our simulations reveal the formation of magnetic structures known as Alfvén wings, which transport magnetic energy away from the planet. When the dayside mass-loss rate $\dot{M}_d$ of the planet lies below a threshold $\dot{ M }_0$ defined by pressure balance between the planetary and stellar winds ($\dot{ M }_d < \dot{ M }_0$), the maximal power delivered to the star matches predictions from the Alfvén wing model. For higher escape rates, the planetary outflow opens additional magnetic flux, and the SPMI power increases proportionally with $(\dot{ M }_d/\dot{ M }_0)^{1/2}$. Applying this scaling law to the HD18973 system, we find that a $50$ G planet could reproduce the observed power if $\dot{ M }_d \sim 5 \times10^{11}\,$g/s. Although this signal likely represents only a fraction of the total power, additional mechanisms could amplify the energy budget. These results show that photoevaporating exoplanets in sub-Alfvénic orbits constitute promising targets for SPMI observations.
\end{abstract}




\section{Introduction}
Many of the exoplanets discovered so far orbit at close distances from their host stars. At such short distances (less than 0.1 au), the planets can experience strong interactions with their host stars through gravity, radiation, stellar winds and magnetic fields \citep{vidotto25}. The observational signatures of these interactions can be used to derive atmospheric escape properties \citep[e.g.][]{bourrier13,zhang22} and stellar wind conditions \citep[e.g.][]{kislyakova14,vidotto23}. Additionally, star-planet magnetic interactions (SPMI) are of special interest because they provide one of the few ways to constrain exoplanetary magnetism \citep[e.g.][]{cauley19,turner21}, which remains largely unknown.

Signatures of SPMI can be observed both at the planet and at the host star. At the planet, the interaction between the stellar wind and the planetary magnetosphere can induce auroral emission from the planet's magnetic poles \citep{callingham24}. In the Solar System, auroral radio emission has been observed in the Earth and the gas giant planets \citep{zarka01}, and it is expected that exoplanets emit at radio frequencies in a similar way \citep{zarka07}. 
So far, a tentative detection of radio emission from an exoplanet has been announced \citep{turner21}. Recently, \citet{tasse26} also found radio emission consistent with SPMI in several systems.

The effects of SPMI can also be observed at the host star when the stellar magnetic energy dominates over the stellar wind kinetic energy at the orbital location of the planet. In this scenario, known as sub-Alfvénic regime, the planet and the star can be magnetically connected, allowing disturbances of the stellar wind generated by the planet to propagate along flux tubes towards the star in the form of Alfvén waves. When the associated energy flux reaches the stellar surface, it can cause enhanced chromospheric activity \citep{shkolnik03}, radio bursts \citep{pineda23}, and flares synchronized with the orbital period of the planet \citep{ilin24,whitsett25}. 
 
In the Solar System, analogous sub-Alfvénic interactions occur between Jupiter and the Galilean satellites Io, Europa, Callisto and Ganymede. The auroral footprints of the moons have been detected in the UV \citep{clarke98,Bhattacharyya18}, in the infrared \citep{mura18}, and by in-situ magnetic field measurements \citep{acuna81}. Even though Solar System planets orbit too far away from the Sun to experience sustained sub-Alfvénic interactions, electromagnetic star-planet coupling might be common in close-in exoplanetary systems \citep{Saur13,atkinson24,strugarek24}. Supporting evidence includes the detection of enhanced Ca emission correlated with the orbital period of the planet in several hot Jupiter systems \citep{shkolnik05,shkolnik08,cauley18,cauley19}, and the detection of coherent radio bursts from the M dwarf YZ Ceti that are associated with the orbital phase of its close-in terrestrial planet \citep{pineda23}. More recently, \cite{ilin25} reported flares on the G-dwarf HIP 67522 that appear to be induced by its innermost planet.    

By comparing analytical theories of SPMI with the energy output of these observations, it is possible, in principle, to place constrains on the magnetic field strength of the planet \citep{lanza18,cauley19,turner21,klein22}. Much of the theoretical understanding of SPMI is inspired by the moon-magnetosphere interactions observed in the Jovian system.
The current understanding is that the companion (moon or planet) is an obstacle in the ambient plasma flow that disturbs the host's magnetic field along their orbit. The perturbations propagate along magnetic field lines towards the host in the form of Alfvén waves, and dissipate part of their energy into the host. This theory of SPMI is known as the Alfvén wing model \citep{neubauer1980,southwood80,Saur13}. The Alfvén wing model predicts energy fluxes consistent with the brightness observed in the Jovian moon-generated aurorae \citep{Saur13}, and has been validated by several numerical simulations \citep[e.g.][]{ridley07,strugarek15,chen25}. However, when applied to extrasolar planets, the Alfvén wing model usually predicts powers about an order of magnitude lower than the ones suggested from observations \citep{cauley19,ilin25}. Alternatively, force-free or stressed loop models can produce much larger powers \citep{lanza13}, albeit those scenarios have not been reproduced by numerical simulations so far. 

These discrepancies show that our understanding of how much SPMI power is generated in close-in gaseous planets is still incomplete. To assess the energetics of SPMI, it might be important to consider the differences between moon-magnetospheric interactions and planet-star systems. Unlike moons, many close-in exoplanets are extended, gaseous bodies subject to intense stellar irradiation that drives atmospheric escape. Atmospheric evaporation has been probed through observations of transiting exoplanets for several systems \citep[e.g.][]{vidalmadjar2003,lecavalierdesetangs2010,orell-miqel24}, and some of these planets experience significant magnetic interactions as well. A notable example is the hot Jupiter HD 189733 b, for which evidence exists both for an electromagnetic coupling to the host star \citep{cauley18,cauley19} and for strong atmospheric evaporation \citep{lecavalierdesetangs2010,bourrier13}.  

While numerical simulations have explored atmospheric escape and SPMI separately, it is not known how these two forms of star-planet interaction influence one another. In particular, the presence of an extended planetary atmosphere could alter the effective size of the planetary obstacle, potentially modifying the power released in the magnetic interaction. Here, we address this question by performing three-dimensional radiation magneto-hydrodynamic (MHD) simulations that self-consistently model both planetary evaporation and magnetic star-planet interactions. We present our numerical model in Section \ref{sec:2}, where we also introduce SPMI and how to estimate its energetics directly from the simulations. Section \ref{sec:3} discusses how stellar irradiation affects SPMI, while Section \ref{sec:4} and Section \ref{sec:5} present our results from different planetary magnetic fields and obliquities. We discuss and give some concluding remarks in Section \ref{sec:6}.

\section{3D modeling of the interaction between a magnetised planet and a sub-alfvenic stellar wind }\label{sec:2}
\subsection{A grid of MHD simulations}\label{sec:2.1}

We use 3D radiation-MHD simulations to quantify the magnetic energy fluxes that arise when a magnetised stellar wind interacts with a magnetised hot Jupiter undergoing atmospheric photoevaporation. Our model solves the ideal MHD equations in a reference frame co-rotating with the planet, which is placed at the origin of the computational domain. 
This model has been employed to study atmospheric escape under super-Alfvénic interactions \citep{Carolan21}, sub-Alfvénic interactions \citep{presa24}, and during coronal mass ejections and flares \citep{Hazra22,hazra25}. The planetary evaporation is simulated by self-consistently solving heating, cooling, ionization and recombination of neutral and ionised hydrogen. The planetary atmosphere is initially neutral, and becomes progressively ionised as it escapes from the planet. Full model details are available in the cited works.

Here, we consider a star-planet configuration similar to HD 209458, including a stellar wind with a radial magnetic field and a dipolar planetary magnetic field. The basic simulation setup and the stellar and planetary parameters are identical to those in \cite{presa24} (see Table 1 for a summary of the system parameters). We consider the models presented in \cite{presa24} that have planetary magnetic field strengths $\geq 2.5 \,$G, so that the planet forms a magnetosphere and the atmospheric escape is magnetically controlled. 
We run an additional series of simulations varying the stellar irradiation $F_{\rm XUV}$ between $250$ and $30000$ erg\,cm$^{-2}$\,s$^{-1}$, resulting in a total of 26 models. 
A summary of the main parameters for each run and the magnetic interaction powers computed here is given in Appendix \ref{appendix:summary_table}. To assess the energetics of the interaction it is important to know the stellar wind conditions at the orbital location of the planet. The wind parameters are listed in Table \ref{tab:wind_parameters}. We discuss the meaning and role of these parameters for the star–planet interaction in the next subsection.

\begin{table}
    \centering
    \caption{Ambient mean stellar wind properties  used in the numerical simulations. Rows list the stellar X-rays and extreme ultraviolet (XUV) flux $F_{\rm XUV}$, magnetic field strength $B_0$, wind speed $u_0$, density $\rho$, Alfvén speed $u_A$, angle between the velocity and magnetic field $\theta$, and Alfvén Mach number $M_A$. All the quantities are given at the orbital location of the planet $a=0.05\,$au, in the planet reference frame.}
\begin{tabular}{lr}
\hline
$F_{\rm XUV}$ (erg\,cm$^{-2}$\,s$^{-1}$) & $2.5\times10^2$ -- $3\times10^4$ \\
$B_0$ (G) & $0.04$ \\
$u_0$ (km\,s$^{-1}$) & $148$ \\
$\rho$ (g\,cm$^{-3}$) & $2\times10^{-19}$ \\
$u_A$ (km\,s$^{-1}$) & $252$ \\
$\theta$ ($^\circ$) & $36$ \\
$M_A$ & $0.59$ \\
\hline
\end{tabular}
    \label{tab:wind_parameters}
\end{table}

\subsection{Structure of star-planet magnetic interactions}

Planets are surrounded by a flow of magnetised plasma that is emitted from the host star, the stellar wind. As a planet moves through this flow it behaves as an obstacle, perturbing the stellar wind and exciting MHD waves that carry electromagnetic energy fluxes. Star-planet magnetic interactions occur when a fraction of this energy is channeled back towards the host star along magnetic field lines. Such electromagnetic coupling requires the relative velocity between the planet's orbital motion and the stellar wind speed to be sub-Alfvénic, a regime where the Alfvénic Mach number $M_A$ is below unity:  
\begin{equation}\label{eq:mach}
    M_A = \mathbf{u_0}/\mathbf{u}_A <1.
\end{equation}
Here, $\mathbf{u_0} = \mathbf{u}_{\rm sw} - \mathbf{u}_{K} $ represents the relative velocity between the stellar wind velocity $\mathbf{u}_{\rm sw}$ and the planet's Keplerian motion $\mathbf{u}_{K}$, while $\mathbf{u}_{A} = \mathbf{B_{0}} / \sqrt{4\pi\rho}$ is the Alfvén speed of the stellar wind (here, $\rho$ and $\mathbf{B_0}$ are the ambient stellar wind density and magnetic field, respectively). When condition \eqref{eq:mach} is fulfilled, there may be a propagation of electromagnetic energy towards the host star, either through Alfvén waves \citep{neubauer1980,Saur13} and/or by release of energy stored in stressed magnetic loops \citep{lanza09,lanza13}. 

\begin{figure}
    \centering
    \includegraphics[width=0.49\textwidth]{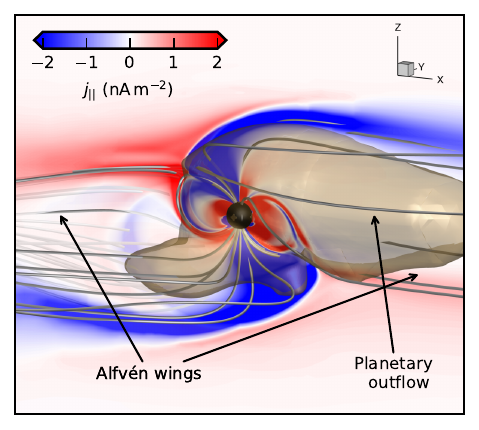}
    \caption{Three dimensional view of the field-aligned current density for a planet (central black sphere) with a polar dipolar field strength of 10$\,$G, irradiated by an incident stellar XUV flux of 500$\,$erg/cm$^2$/s. The dipole axis is oriented at $\Theta_M=90^\circ$ relative to the incident stellar wind magnetic field. Positive and negative currents are color‑coded in red and blue, respectively. Two distinct Alfvén wings, bounded by regions of strong parallel current density, are visible: one wing extends toward the negative-$X$ direction and remains magnetically connected to the stellar field. The other wing extends along the positive-$X$ direction, away from the star.
    The brown volume rendering represents the photo-evaporating planetary atmosphere, while the streamlines trace the magnetic field of the planet. }
    \label{fig:FAC}
\end{figure}

We show the magnetic configuration that results from our simulations in Figure \ref{fig:FAC}, where the grey streamlines indicate the planetary magnetic field lines, and the planet is shown as the central black sphere. The magnetospheric boundary and the open-field regions are bounded by strong field-aligned current densities, $\mathbf{j}_{||}=(\mathbf{j}\cdot\mathbf{B})/|\mathbf{B}|$, shown color-coded in the figure. These  currents are also strong in the transitional region between the stellar wind and the magnetic flux tubes where Alfvén wing propagate. Alfvén waves are transversal, non-compressional waves that propagate with a group velocity that is parallel or anti-parallel to the background magnetic field \citep{goedbloed19}. This means that they carry perturbations in magnetic field perpendicular to the background field $\delta B_\perp$ over large distances \citep{saur18}.

The superposition of Alfvén wave-packages form a pair of field-aligned current systems known as Alfvén wings \citep{drell65,neubauer1998}. The direction of the Alfvén wings is given by the Elsässer variables \citep{elsasser50}, also known as Alfvén characteristics:
\begin{equation}\label{eq:characteristics}
    \mathbf{c}^{\pm}_A=\mathbf{u}_0 \pm \mathbf{u}_A.
\end{equation}
Equation \eqref{eq:characteristics} indicates that the perturbations travel along magnetic field lines while being displaced by the background plasma \citep{neubauer1998,Saur13,strugarek15}. The Alfvén wings that develop in our simulations are indicated by arrows in Figure \ref{fig:FAC}. Due to the positive radial topology of the stellar field, only the ``$-$'' solution points towards the star (located in the negative-$X$ direction of the grid). Since we are interested in the power transported toward the host star, we limit our analysis to this wing. In the next subsection, we describe how to compute the associated power.

\subsection{Energetics of star-planet magnetic interactions}\label{sec:energetics_calculation}
The interaction between the magnetised stellar wind and the planet perturbs the flow and redirects part of its electromagnetic energy into the Alfvén wings. The Poynting flux flowing through the wings is found by projecting the stellar wind Poynting vector $\mathbf{S}$ along the characteristics given by Equation \eqref{eq:characteristics} \citep{strugarek15}:
\begin{equation}\label{eq:S_aw,tot}
S_{\rm AW,tot} = \mathbf{S} \cdot \frac{\mathbf{c}^\pm_A}{|\mathbf{c}^\pm_A|} =-\frac{(\mathbf{u} \times \mathbf{B}) \times \mathbf{B}}{4\pi} \cdot \frac{\mathbf{c}^\pm_A}{|\mathbf{c}^\pm_A|}.   
\end{equation}
Here, we wrote the electric field as $\mathbf{E}= - (\mathbf{u} \times \mathbf{B})/c$ in the ideal MHD approximation, where $c$ is the speed of light. 

The Poynting flux \eqref{eq:S_aw,tot} depends on the reference frame of the velocity $\mathbf{u}$. Since we are interested in the energy flux deposited into the star, we follow \citet{Saur13,strugarek15} and describe the stellar wind in the reference frame of the central star. To isolate the direct contribution from the planet, we subtracted the Poynting flux of the unperturbed stellar wind $S_0=-\frac{(\mathbf{u}_0 \times \mathbf{B}_0) \times \mathbf{B}_0}{4\pi} \cdot \frac{\mathbf{c}^\pm_A}{|\mathbf{c}^\pm_A|}$ from the total Poynting flux:
\begin{equation}\label{eq:S_aw}
    S_{\rm AW} = S_{\rm AW,tot} - S_0.
\end{equation}
Here, $S_{\rm AW}$ is the net Poynting flux  due the star-planet magnetic interaction.
In practice, to set the background values, we used a reference model with low planetary field $B_p=0.1\,$G. In this model, the planet does not form a significant magnetosphere,
and we verified that the Alfvénic perturbation generated by the planet is too weak to be numerically resolved at the analysis plane used for the power calculations. Therefore, we can consider the $B_p=0.1\,$G model as a background reference case where the stellar wind is mostly unperturbed far away from the planet. 

The total power carried by the Alfvén wing is then found by integrating $S_{\rm AW}$ through an area $\mathcal{A}$ normal to the Alfvén wing, at a certain distance from the planet. We integrate only in the subset of points $\mathcal{A_+}$ within that plane where $\mathbf{u}\cdot\mathbf{c_A^-}>0$:
\begin{equation}\label{eq:power}
        \mathcal{P}_{\rm AW} = \int_{\mathcal{A}_+} {S}_{\rm AW} \, \textrm{d}\mathcal{A}_+ \quad  \mathcal{A}_+ = \{\mathbf{x}\in \mathcal{A}: \mathbf{u}(\mathbf{x})\cdot\mathbf{c^-_A} >0\}. 
\end{equation}
We evaluate the wing cross-section on the $YZ$ plane at some distance $X$ from the planet  (i.e. on planes normal to the star–planet line). This analysis plane is approximately perpendicular to the characteristics \eqref{eq:characteristics}.
In principle, $\mathcal{P}_{\rm AW}$ could be computed at any distance $X$ from the planet provided that contributions from the fast and slow MHD modes are negligible \citep{Saur13}. In numerical simulations, however, Alfvén wings typically lose small amounts of power with distance due to numerical dissipation and coarsening of the computational mesh \citep[e.g.][]{ridley07,fisher22}. We observe the same effect in our models, particularly at large separations from the planet where the wing structure becomes under-resolved. To balance these effects, we therefore restrict our analysis to planar cuts taken at $-18\,R_p \leq X \leq -26\,R_p$. We show the computed $\mathcal{P}_{\rm AW}$ at different distances in Appendix \ref{appendix:power_distance}.
\begin{figure*}
    \centering
    \includegraphics[width=0.999\textwidth]{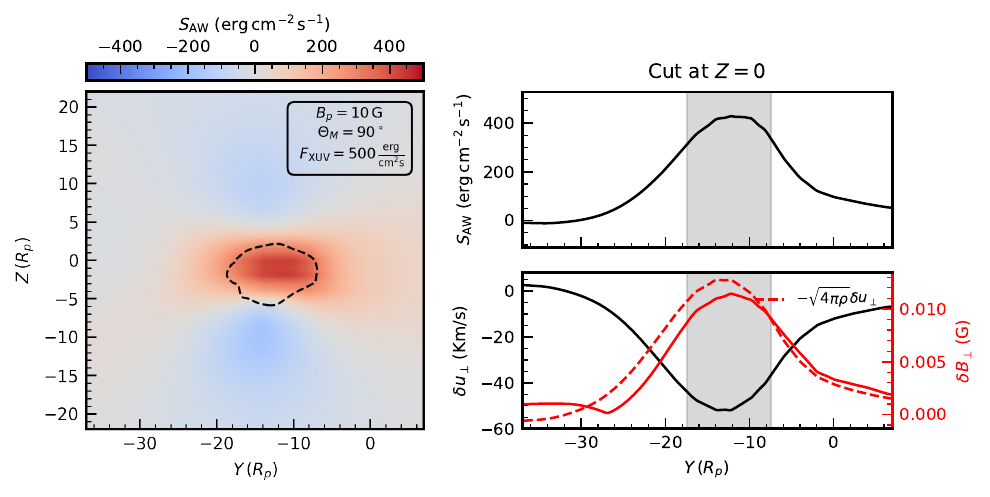}
    \caption{Left: cut along the Alfvén wing at $X=-24 R_p$ for the model with $B_p=10$\,G and $\Theta_M=90^\circ$ shown in Figure \ref{fig:FAC}. 
    Colors indicate the Poynting flux. Black contours represent the boundary of the inner Alfvén wing, identified as the cross-section of the flux tube connecting the star and the planet. Right: properties of the Alfvén wing through the $Z=0$ line along the $X=-24 R_p$ plane. The top panel represents the Poynting flux, the bottom panel represents the Alfvénic perturbations in velocity (solid black line) and magnetic field (solid red line). The dashed red line indicates the analytical prediction of linear MHD theory for an Alfvén wave. }
    \label{fig:Poynting_plane}
\end{figure*}
Figure \ref{fig:Poynting_plane} illustrates the properties of the $\mathbf{c_A^-}$ wing for the $10\,$G planet shown in Figure \ref{fig:FAC}. The left panel shows the Poynting flux $S_{\rm AW}$ on the cut at $X=-24\,R_p$, where 
the dashed black line delimits the cross-section of the flux tube connecting the star and the planet. We found this cross-section by directly tracing the magnetic field lines that stream from the planet. 
The flux tube forms the inner part of the Alfvén wing, and carries the strongest Poynting flux generated by the planetary obstacle. 
This is also visible in the right panels of Figure \ref{fig:Poynting_plane}, where we show several cuts along the $Z=0$ line in the $X=-24\,R_p$ plane. 
The top panel shows $S_{\rm AW}$, while the bottom panel displays 
the perturbations in velocity $\delta u_\perp$ (solid black line) and magnetic field $\delta B_\perp$ (solid red line) perpendicular to the field along the same cut. These perturbations are calculated as
\begin{flalign}\label{eq:perturbations}
    \delta u_\perp = (\mathbf{u}-\mathbf{u_0})\cdot \mathbf{\hat{e}_{\perp \, B_0}} ,\\
    \delta B_\perp = (\mathbf{B}-\mathbf{B_0})\cdot \mathbf{\hat{e}_{\perp \, B_0}} ,
\end{flalign}
where $\mathbf{\hat{e}_{\perp \, B_0}}$ is the unit vector perpendicular to the background magnetic field.
The strongest perturbations match the location of the flux tube, represented here with a gray shade. In the bottom panel, the dashed red curve is the $\delta B_\perp$ predicted by linear MHD wave theory given the measured $\delta u_\perp$. The close match between the simulated and the theoretical $\delta B_\perp$ confirms the Alfvénic nature of the perturbations, and that the influence of other wave modes is negligible. 

Outside the inner wing, the Poynting flux, the velocity and magnetic field perturbations do not return sharply to zero. Instead, they decrease gradually because the flux tube itself also acts as an obstacle to the stellar wind \citep{chane12,strugarek15,Paul25}. To account for the total power transmitted to the star we therefore integrate the net Poynting flux $S_{\rm AW}$ over the region where $S_{\rm AW}>0$. We note that $S_{\rm AW}$ takes small negative numbers in some regions outside the Alfvén wing (see the left panel in Figure \ref{fig:Poynting_plane}), this means that the perturbed Poynting flux is smaller than the background stellar wind Poynting flux in those areas. As argued in \cite{Paul25}, this can be understood from local energy conservation: the planetary obstacle channels a fraction of the stellar wind energy towards the star; this energy must be supplied from the surrounding stellar wind.  

\section{Effect of atmospheric escape on Alfvén wings and SPMI power}\label{sec:3}
In this section, we examine how such planetary outflows can affect the Alfvén
wing that links a planet to its host star. First, we give a semi-analytical description of the problem in Section \ref{sec:3.1}. Then, we conduct a series of MHD simulations with
different stellar irradiation conditions in section \ref{sec:3.2}. 

\subsection{Semi-analytical description of the interaction between planetary
outflows and Alfvén wings}\label{sec:3.1}
Close-in exoplanets receive a substantial amount of stellar irradiation, which leads to photoevaporation of their atmospheres. To affect the Alfvén wing that links a planet to its host star, the planetary outflow must be launched, at least in part, on the dayside. However, planetary outflows do not expand freely in the vacuum; they are bound by the
surrounding stellar wind. If the stellar wind has a sufficiently high pressure, it can reduce the dayside flow into a subsonic breeze, or even suppress it \citep{murray-clay09,vidotto20}. 

To ensure that the dayside mass loss is not shut off completely, we require that the ram pressure of the stellar wind does not exceed the ram pressure of the outflow close to the planet. Approximating mass-loss rates under spherical symmetry, this condition can be written as \citep{owen14}:
\begin{equation}\label{eq:condition_dayside}
     \Pi = \frac{\dot{M}_\star u_{\rm sw}(1+x)^2R^2_{p}}{\dot{M}u_{\rm pw}a_p^2} < 1.
\end{equation}
Here, $\dot{M_\star}$ and $\dot{M}$ are the stellar and planetary mass-loss rates, respectively. $a_p$ is the orbital radius of the planet, $u_{\rm sw}$ is the stellar wind velocity, and $u_{\rm pw}$ is the velocity of the planetary outflow.
$R_{\rm XUV}=(1+x)R_p$ is the radius where XUV photons are absorbed, which we consider to be the outflow launching radius. This radius can be estimated using equation 4 in \citet{salz16}. For the hot Jupiter HD209458b and $F_{\rm XUV}=10^2-10^3\,$erg$\,$cm$^{-2}\,$s$^{-1}$, this results in $R_{\rm XUV}=1.1-1.2\,R_p$. This implies that, for hot Jupiters, $R_{\rm XUV}$ is typically very close to the planetary radius, and will be smaller than the sonic radius $R_s$ ($x\ll1$ and $R_{\rm XUV} < R_s$). Therefore, Equation \eqref{eq:condition_dayside} is a weaker version of the non-confinement condition $P_{\rm ram,sw}(a_p) < P_{\rm ram,pw}(R_s)$ given in \citet{vidotto20}. In the latter case, a subsonic breeze could still be possible, which would still affect the Alfvén wing. By setting $\Pi=1$, we can find the minimum mass-loss rate $\dot{M}_{0}$ that would support a dayside outflow:
\begin{equation}\label{eq:M_lower}
     \dot{M}_{0} = \frac{\dot{M}_\star u_{\rm sw}(1+x)^2R^2_{p}}{u_{\rm pw}a_p^2}.
\end{equation}

Additionally, for a strong planetary outflow and/or weak planetary magnetic field, the outflow blows open the field radially close to the planet. In this scenario, the outflow is no longer magnetically controlled and the planet does not form a magnetosphere \citep{presa24}. 
We define a second mass loss threshold $\dot{M}_1$ by considering that the plasma is magnetically controlled when $M_A<1$, now evaluating the Mach number with the planetary wind speed and density (i.e. $u=u_{\rm pw}$, $\rho=\rho_{\rm pw}$). An analogous condition can be derived by comparing the ram pressure of the outflow against the magnetic pressure of the planetary field \citep{owen14}.
We evaluate the condition $M_A<1$ at the radius $R_B$ from the planet where the stellar and planetary magnetic field strengths are equal, $B_{p}(R_B) = B_\star(R_B)$:
\begin{equation}\label{eq:Mdot_1_1}
    M_A = \frac{\sqrt{4\pi\rho_{\rm pw}}u_{\rm pw}}{B_\star(R_B)}<1.
\end{equation}
For a dipolar planetary field with surface field strength $B_{p,0}$, $B_p(r) = B_{p,0}(R_p / r)^3$ and $R_B$ is given by
\begin{equation}\label{eq:R_B}
     \frac{R_B}{R_p} = \Big(\frac{B_{p,0}}{B_\star(a_p)}\Big)^{1/3},
\end{equation}
where we assumed that $R_B$ is much smaller than the orbital distance of the planet $a_p$, so that $B_\star(R_B) \approx B_\star(a_p)$. Combining Equations \eqref{eq:Mdot_1_1} and \eqref{eq:R_B}, and assuming a spherically symmetric outflow $\dot{M}=4\pi R^2_B \rho_{\rm pw} u _{\rm pw}$, we can find the maximum planetary mass-loss rate that allows a magnetically controlled atmospheric escape:
\begin{equation}\label{eq:M_1}
     \dot{M}_{\rm 1} = \frac{B_{p,0}^{2/3}B_\star(a_p)^{4/3}R_p^2}{u_{\rm pw}}.
 \end{equation}

When the dayside mass-loss rate $\dot{M}_d \in (\dot{M_0}, \dot{M}_1)$, the flux tube connecting the star and the planet will be partially loaded with planetary material close to the planet. This material adds an additional contribution to the lateral pressure balance, increasing the Alfvén wing cross-section. Additionally, the planetary plasma motions can also amplify the magnetic field by stretching and stressing the field lines. Both effects increase the open magnetic flux available for the star–planet connection $\Phi_{\rm conn}$.
  
To write the Alfvén wing power in terms of $\Phi_{\rm conn}$ one needs to calculate Equation \eqref{eq:power}. \citet{Saur13} showed that, for small Mach numbers, $\mathcal{P}_{\rm AW}$ can be written in closed form as
\begin{equation}\label{eq:power_aw_saur}
    \mathcal{P}_{\rm AW, th} = \alpha^2 \left( \frac{M_Au_0 B^2_0\sin^2\theta}{4\pi} \right)  2 \pi R^2_{\rm eff}.
\end{equation}
The term in parentheses in Equation \eqref{eq:power_aw_saur} represents the Poynting flux carried away along the Alfvén wing towards the star. Here, $u_0$ is the relative speed between the stellar wind and the planet, $B_0$ is the unperturbed stellar wind magnetic field strength, $\theta$ represents the relative angle between $u_0$ and $B_0$, $M_A$ is the Mach number given by Equation \eqref{eq:mach}, and $\alpha$ is the strength of the sub-Alfvénic interaction. This factor is defined as
\begin{equation}\label{eq:alpha_saur}
    \alpha \simeq 1 - \frac{u}{u_0}, 
\end{equation}
and it describes how strongly the stellar wind velocity $u$ is reduced in the vecinity of the planet or inside the wings.

The flux in Equation \eqref{eq:power_aw_saur} is then multiplied by the effective area of the interaction $A_{\rm eff}=2 \pi R_{\rm eff}^2$. \citet{Saur13} estimate the effective size of the obstacle $R_{\rm eff}$ as the average width of the flux tube connecting the star and the planet, given by 
\begin{equation}\label{eq:Reff:saur}
    R_{\rm eff}/R_m = \sqrt{3 \cos\left({\frac{\Theta_M}{2}}\right)}.  
\end{equation}
Here, $\Theta_M$ is the relative orientation between the stellar wind and planetary magnetic moment, and $R_m$ is the magnetospheric size of the planet. The effective area of the interaction is a factor of 2 larger than the tube's cross-section $\pi R_{\rm eff}^2$, as it accounts for the energy flux transmitted both inside and outside the flux tube itself \citep{Saur13}. 

By approximating the magnetic flux of the flux tube connecting the star and the planet as $\Phi_{\rm conn} \simeq B_0 \pi R_{\rm eff}^2$, we can rewrite Equation \eqref{eq:power_aw_saur} as
\begin{equation}\label{eq:power_saur_flux}
    \mathcal{P}_{\rm AW, th} = \alpha^2\frac{u^2_0\sin^2\theta}{2\pi} \sqrt{4\pi\rho} \Phi_{\rm conn}.
\end{equation}

To account for the effects of atmospheric evaporation in the Alfvén wing power, one needs to know the dependence between the planet dayside mass-loss rate $\dot{M}_d$ and the magnetic flux $\Phi_{\rm conn}$. We note here that $\Phi_{\rm conn}$ will be a fraction of the open magnetic flux of the planet $\Phi_{\rm open}$. In stellar wind theory, the spherically symmetric model developed by \citet{WeberDavies67} predicts a relation $\dot{M_\star} \propto \Phi_{\rm open}$. Similarly, numerical simulations find a tight correlation of the form $\dot{M_\star} \propto \Phi_{\rm open}^p$, with a power-law index $p$ taking values between $0.5$ and $1$ \citep[e.g.][]{vidotto14,Evensberget22}. Considering these results, we write $\Phi_{\rm conn}$ as
\begin{equation}\label{eq:Phi_open_powerlaw}
    \Big( \frac{\Phi_{\rm conn}}{\Phi_{0}}\Big)=\Big(\frac{\dot{M}_d}{\dot{M}_0}\Big)^\gamma,
\end{equation}
where we defined $\Phi_0 :=\Phi_{\rm conn}(\dot{M}_d\leq \dot{M_0})$ and $\gamma$ is a free parameter that can be constrained using simulations of star-planet magnetic interactions coupled with atmospheric escape. Using Equation \eqref{eq:Phi_open_powerlaw}, the Alfvén wing power can be rewritten as 
\begin{equation}\label{eq:P_AW_Mdot}
    \mathcal{P}_{\rm AW,esc}=\mathcal{P}_{\rm AW, th}\Big(\frac{\dot{M}_d}{\dot{M}_0}\Big)^\gamma \;\; \text{for $\dot{M}_d \in (\dot{M_0}, \dot{M}_1)$}.
\end{equation}

Last, we note that, for $\dot{M_d} \geq \dot{M}_1$, the planet does not form a magnetosphere, so we expect the power to be reduced to the theoretical \citet{Saur13} value given by Equation \eqref{eq:power_aw_saur} for an unmagnetised planet, i.e. $R_{\rm eff} = R_p$.

 \subsection{MHD simulations of Alfvén wings under different
planetary mass-loss rates}\label{sec:3.2}
To analyse numerically how atmospheric escape modifies the Alfvén wing, we perform a suite of MHD simulations following the model described in section \ref{sec:2.1}. To isolate the effect of atmospheric escape, we fix the planetary magnetic field strength to $B_p=10\,$G and set its relative orientation with respect to the stellar-wind magnetic field to $\Theta_M=90^\circ$ throughout the simulations. The stellar wind properties are adopted from \citet{presa24}, and are also kept constant in all the simulations. The main stellar wind properties at the orbital location of the planet are listed in Table~\ref{tab:wind_parameters}. We use a polytropic 1D model to fix the stellar wind velocity and temperature at the boundary of the 3D simulation, similar to \citet{carolan21sw,Carolan21}. Here, we considered a polytropic index of $\gamma=1.0001$, resulting in a stellar wind temperature profile that is essentially isothermal. The stellar wind mass-loss rate and magnetic field were chosen to represent an active solar-like star, while ensuring that the wind remains sub-Alfvénic at the planet’s orbital distance. 
We then vary the incident XUV flux $F_{\rm XUV}$ from $250$ to $30000\,$erg$\,$cm$^{-2}\,$s$^{-1}$, effectively changing the structure and mass-loss rate of the planetary outflow. The flux values used in the simulations are listed in Table \ref{tab:summary}.

\begin{figure*}
    \centering
    \includegraphics{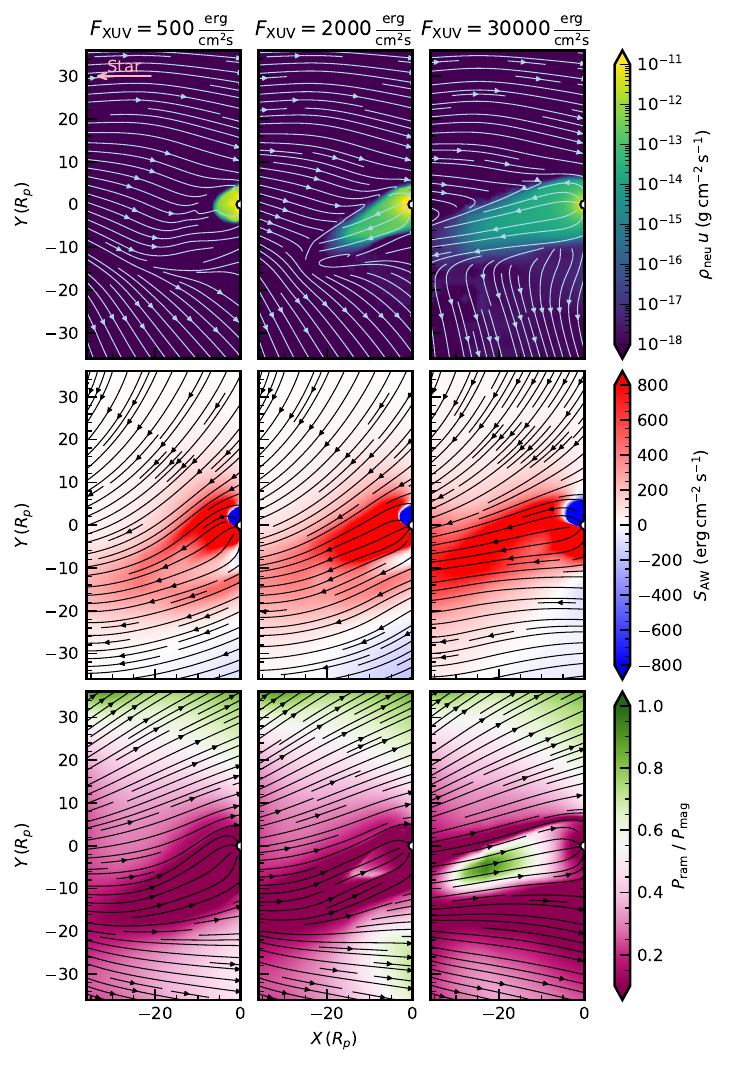}
    \caption{Cut at the $Z=-2\,R_p$ (right below the orbital plane) plane for a $B_p=10\,$G, $\Theta_M=90^\circ$ planet under three representative stellar irradiation conditions. The planet is centered at the origin, and the star is located in the negative-$X$ direction, outside the computational grid. Each column shows a different stellar XUV flux, corresponding to distinct atmospheric escape scenarios: no dayside outflow ($F_{\rm XUV}=500\,$erg$\,$cm$^{-2}$$\,$s$^{-1}$), a dayside breeze ($F_{\rm XUV}=1000\,$erg$\,$cm$^{-2}$$\,$s$^{-1}$), and a dayside wind ($F_{\rm XUV}=30000\,$erg$\,$cm$^{-2}$$\,$s$^{-1}$). The top row represents the flux of planetary neutral material, while the streamlines trace the velocity flow. The middle row shows the Alfvén wing Poynting flux $S_{\rm AW}$, and the streamlines correspond to the Alfvén characteristics $\mathbf{u}-\mathbf{u_A}$. The bottom row shows the ratio between the ram and magnetic pressures, $P_{\rm ram}$ and $P_{\rm mag}$, respectively. The black streamlines represent the magnetic field lines.}
   \label{fig:XUV_plane}
\end{figure*}

The top panel in Figure \ref{fig:XUV_plane} shows the dayside mass flux of planetary neutral material for three different stellar irradiation conditions, with streamlines indicating the flow velocity. For the case shown in the left column ($F_{\rm XUV}=500\,$erg$\,$cm$^{-2}\,$s$^{-1}$), the planetary material is confined by the stellar wind pressure in the vicinity of the planet. Here, the planetary wind mainly consists of a single polar stream directed away from the star along the open planetary field lines that emerge from the northern magnetic pole (see Figure \ref{fig:FAC} and \citealt{presa24}). 
As the incident XUV flux increases, a dayside outflow structure develops progressively. For an intermediate irradiation ($F_{\rm XUV}=2\times10^3\,$erg$\,$cm$^{-2}\,$s$^{-1}$, middle column)  a weak breeze steams from the dayide of the planet, which stagnates at some distance from the planet. At higher XUV fluxes, the dayside breeze transitions into a wind. The planetary wind travels downstream towards the simulation boundary, mainly channeled along the stellar–planetary flux tube. This structure can be seen in the top-right panel of Figure \ref{fig:XUV_plane}, corresponding to $F_{\rm XUV}=3\times10^4\,$erg$\,$cm$^{-2}\,$s$^{-1}$.

From Section \ref{sec:3.1}, we expect an increase in the Alfvén wing power when the flux tube linking the planet and the star is partially filled with planetary material. This behaviour is observed in the Poynting flux $S_{\rm AW}$ distribution, shown in the middle row panels of Figure \ref{fig:XUV_plane}. In all our models, $S_{\rm AW}$ flux is always directed toward the host star, as the characteristics $\mathbf{c_A^-}$ (black streamlines) point towards the star.  
We observe broader Alfvén wings in the models that exhibit dayside mass loss. This broadening results from the expansion of the flux tube caused by the planetary outflow. Additionally, regions filled with planetary material carry stronger Poynting fluxes. In our simulations, the planetary outflow also perturbs the magnetic field within the wing, leading to an increase in the Poynting flux. This is clearly seen in the lower panels of Figure \ref{fig:XUV_plane}, where we show the ratio of ram pressure $P_{\rm ram}$ to magnetic pressure $P_{\rm mag}$ and the magnetic field streamlines. In the $F_{\rm XUV}=3\times10^4\,$erg$\,$cm$^{-2}\,$s$^{-1}$ model, the ram pressure of the escaping planetary flow is comparable to the ambient magnetic pressure in the regions filled by planetary material. Under these conditions, the planetary outflow is able to significantly reduce the curvature of the connecting field lines (i.e., stretch them), which produces a straighter flux tube and a  lateral displacement of the Alfvén wing compared with the low-irradiation cases. 

\begin{figure}
    \centering
    \includegraphics[width=0.49\textwidth]{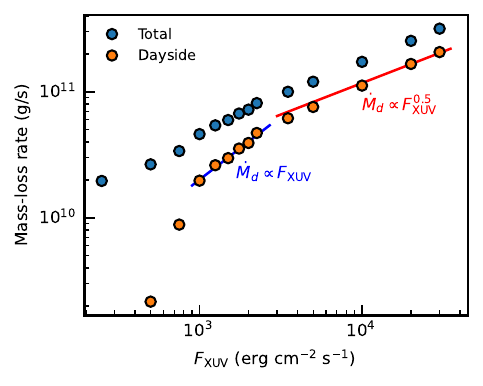}
    \caption{Planetary mass-loss rate $\dot{M}$ dependence on incident stellar XUV flux $F_{\rm XUV}$ for a $B_p=10\,$G, $\Theta_M=90^\circ$ planet. The blue circles represent the total escape rates derived from our simulations, and the orange circles show the integrated mass flux directed towards the star at $5\,R_p$ from the planet. The blue line indicates the theoretical energy-limited mass-loss rate calculated using Equation \eqref{eq:mdot_el} with $x=0.1$ and $\eta=0.3$. The red line indicates the theoretical recombination-limited mass-loss rate calculated using Equation \eqref{eq:mdot_rl} with $A=3$.}
    \label{fig:mass_loss_Fxuv}
\end{figure}

To analyse the effects of atmospheric escape on the Alfvén wing power, we first need to compute the planetary mass-loss rate associated with each $F_{\rm XUV}$ case. We calculate the mass-loss rate by integrating the mass flux of the evaporating material through a cube of area $A$ centered on the planet: 
\begin{equation}\label{eq:mdot_integral}
    \dot{M}=\oint_A \rho_{\rm pw} \mathbf{u}_{\rm pw} \cdot {d}\mathbf{A},
\end{equation}
where $\rho_{\rm pw}$ and $\mathbf{u}_{\rm pw}$ denote the density and velocity of the planetary material, respectively.
We choose a cube with an edge length of $10\,R_p$, so that the integration is performed just outside the magnetosphere of the planet. The blue circles in Figure \ref{fig:mass_loss_Fxuv} represent the total escape rate from each model as a function of $F_{\rm XUV}$. The mass-loss rate ranges between $2\times10^{10}\,$g/s for $F_{\rm XUV}=250\,$erg$\,$cm$^{-2}$$\,$s$^{-1}$, to about $3\times10^{11}\,$g/s for $F_{\rm XUV}=3 \times 10^4\,$erg$\,$cm$^{-2}$$\,$s$^{-1}$.

However, the downstream Alfvén wing is only affected by the evaporating planetary material that flows towards the star, the "dayside mass-loss rate" $\dot{M}_{d}$. We compute the integrated mass flux directed towards the star by adding the mass lost through the $Z=-5\,R_p$ and $X=-5\,R_p$ planes. These planes trace the planetary material flowing along the pole connected to the star. The integrated mass flux directed towards the star at this distance from the planet is shown in Figure \ref{fig:mass_loss_Fxuv} as orange circles. The fraction of the planetary material directed towards the star at $-5\,R_p$ increases with $F_{\rm XUV}$. At the lowest stellar XUV flux there is almost no planetary material moving towards the star, whereas at the highest irradiation level the dayside outflow accounts for more than 50 \% of the total escaping mass. 

While the Alfvén wing power given by Equation \eqref{eq:P_AW_Mdot} depends on $\dot{M}_d$, the mass-loss rate is not a direct observable, and needs to be derived from analytical models or more detailed simulations. Therefore, it is interesting to compare the mass-loss rate $\dot{M}$ with two limiting cases of XUV-driven atmospheric escape. At low fluxes, the energy gained from photoionization is balanced by gas advection, and the escape is energy-limited. In this regime, the mass-loss rate is given by \citep{watson81,erkaev07}
\begin{equation}\label{eq:mdot_el}
     \dot{M}_{\rm EL} = \eta\frac{F_{\rm XUV} (1+x)^2 (\pi R^3_{\rm p})}{KGM_p},
 \end{equation}
 where $\eta$ is the mass-loss rate efficiency, $x$ is the distance from the surface of the planet to the radius where XUV photons are absorbed, and $K$ is a correction factor to account for stellar tidal effects \citep{erkaev07}: 
\begin{equation}\label{eq:tidal_correction}
     K = 1 - \frac{3R_p}{2R_{\rm Roche}} + \frac{3R^3_p}{R_{\rm Roche}^3},
 \end{equation}
 where $R_{\rm Roche}$ is the Roche radius. At high XUV fluxes, ionization is balanced by radiative cooling and recombination. The recombination-limited escape can be written as \citep{Owen12}
\begin{equation}\label{eq:mdot_rl}
    \dot{M}_{\rm RL} = 4\pi (1+x)^{3/2} \mu m_H R^{3/2}_p (A \alpha_r h\nu)^{-1/2} F_{\rm XUV}^{1/2},
\end{equation}
where $m_H$ is the hydrogen mass, $\mu$ is the mean molecular weight, $A$ is a geometric factor, $\alpha_R=2.6 \times 10^{-13}\,$cm$^3$\,s$^{-1}$ is the recombination rate of hydrogen at $10^4\,$K, $h \nu = 20\,$eV is the energy of our monochromatic XUV spectrum, and $m_H = 1.67\times10^{-24}\,$g is the mass of the proton.

In our models, we can reproduce the dayside mass-loss rate $\dot{M}_d$ using Equations \eqref{eq:mdot_el}-\eqref{eq:mdot_rl} taking $x=0.1$, $\eta=2/3$, and $A=3$, and assuming that one half of the total mass computed using these equations is lost through the dayside ($\dot{M}_d=\dot{M}/2$). The resulting fits for energy-limited and recombination-limited escape are shown in Figure \ref{fig:mass_loss_Fxuv} with blue and red lines, respectively. The transition between both escape regimes takes place at $F_{\rm XUV}\sim 3 \times 10^3\,$erg$\,$cm$^{-2}$$\,$s$^{-1}$.

\begin{figure*}
    \centering
    \includegraphics[width=0.999\textwidth]{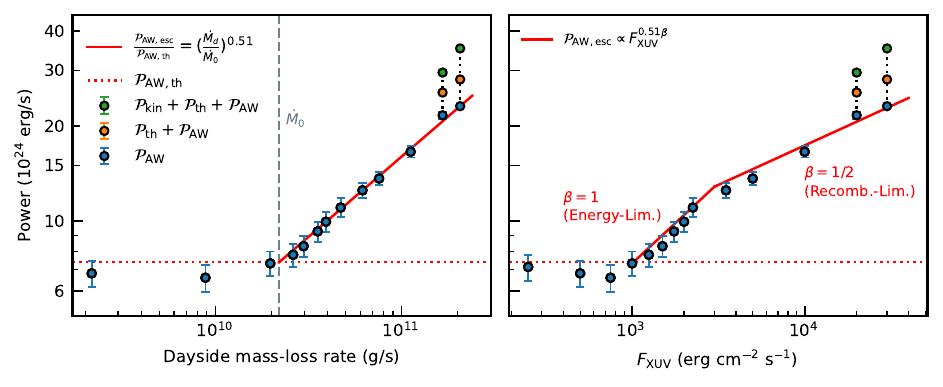}
    \caption{Left: Alfvén wing power as a function of the dayside mass-loss rate for a $10\,$G planet with $\Theta_M=90^\circ$. The blue circles represent the mean integrated Alfvén wing power at several distances from the planet (see text), and the error-bars indicate the $1\sigma$ dispersion from the average. Orange and green circles indicate the (cumulative) contribution of the thermal and kinetic fluxes to the total power. The dashed vertical line indicates the analytical mass-loss rate limit given by Equation \eqref{eq:M_lower}, and the  dotted horizontal line shows the theoretical Alfvén wing power from Equation \eqref{eq:power_aw_saur}.
    The red line is a fit to the semi-analytical form given in Equation \eqref{eq:P_AW_Mdot}. Right: Same as the left panel, but showing the Alfvén wing power as a function of the incident stellar XUV flux. The broken red line corresponds to a fit obtained using the best-fit power-law exponent derived for the left panel, combined with the two mass-loss rate regimes described in Equations \eqref{eq:mdot_el} and \eqref{eq:mdot_rl}.}
    \label{fig:power_mdot_flux}
\end{figure*}

In the left panel of Figure \ref{fig:power_mdot_flux}, we show the Alfvén wing power for each simulation as a function of the computed $\dot{M}_d$ (blue circles). We estimate the power by integrating $S_{\rm AW}$ at planes between $X=-22\,R_p$ and $X=-26\,R_p$, as described in Section \ref{sec:2.1}. We chose these distances to ensure a negligible contribution from other wave modes, while keeping an acceptable grid resolution. 
Our results show that the power remains roughly constant until $\dot{M}_d=2\times10^{10}\,$erg$\,$cm$^{-2}$$\,$s$^{-1}$, and then increases monotonically. This threshold is related to the mass-loss rate limit $\dot{M}_{\rm 0}$ defined in Equation \eqref{eq:M_lower}. As argued in Section \ref{sec:3.1}, this is the minimum mass-loss rate that would support a stream of planetary material through the dayside of the planet. 
It is possible to estimate $\dot{M}_{0}$ using Equation \eqref{eq:M_lower}, with the parameters used in our simulations: $\dot{M}_\star = 2\times10^{-13}\,M_\odot/ \rm yr$, $a_p = 0.05\, \rm au$, and considering $u_{\rm sw}/u_{\rm pw} \approx 10$ (a similar ratio is found in our simulations). Taking again $x=0.1$, we obtain $\dot{M}_{0} =2.2\times10^{10}\,$g/s. We show $\dot{M}_{0}$ in the left panel of Figure \ref{fig:power_mdot_flux} as a gray vertical line. We note that this limit lies very close to the point where the power begins to increase.

Next, we study how the powers computed numerically compare with the analytical Alfvén wing model of \citet{Saur13}. We compute the theoretical power using Equation \eqref{eq:power_aw_saur} with the unperturbed stellar wind parameters given in Table \ref{tab:wind_parameters}. We calculate the strength of the interaction $\alpha$ from Equation \eqref{eq:alpha_saur}, measuring the perturbed velocity $u$ for each model. The effective size of the obstacle is estimated from Equation \eqref{eq:Reff:saur}, where we took $\Theta_M=90^\circ$ and $R_m=4.3\,R_p$ (see Appendix \ref{appendix:magnetosphere} for the estimation of the magnetospheric size). From this calculation, we obtained a theoretical power of $\mathcal{P}_{\rm AW,th}=7.4 \cdot 10^{24}\,$ erg/s (horizontal dotted line in Figure \ref{fig:power_mdot_flux}). We observe a very good agreement between $\mathcal{P}_{\rm AW,th}$ and the numerical results for $\dot{M}_d<\dot{M}_0$.

For stellar fluxes higher than $F_{\rm XUV}=750\,$erg$\,$cm$^{-2}$$\,$s$^{-1}$, the power increases monotonically. In this regime, some planetary material flows from the dayside and can influence the Alfvén wing connected to the star. Therefore, the correction to the Alfven wing power described in Section \ref{sec:3.1} should be considered. To take this into account, we fit the numerically calculated powers to the power law given in Equation \eqref{eq:P_AW_Mdot}. Using the values of $\mathcal{P_{\rm AW,th}}$ and $\dot{M}_0$ computed above, and setting the power-law index $\gamma$ as the free parameter, we obtain a best-fit value of $\gamma = 0.51$. The red solid line in the left panel of Figure \ref{fig:power_mdot_flux} shows the results of this fit, which is in close agreement to the numerical values.

For completeness, we also study the dependence of the power on the incident XUV flux. Using the scalings $\mathcal{P_{\rm AW}}\propto \dot{M}_d^{\gamma}$ and $\dot{M}_d \propto F_{\rm XUV}^{\beta}$, the resulting relation is $\mathcal{P_{\rm AW}}\propto F_{\rm XUV}^{\gamma\beta}$. Because the mass-loss rate regime changes from energy-limited ($\beta=1$) at low irradiation to recombination-limited ($\beta=1/2$) at higher irradiation, the power-flux relation is a broken power law. Based on our mass-loss rate calculations (see Figure \ref{fig:mass_loss_Fxuv}), we adopt $\beta=1$ for $F_{\rm XUV}<3 \times 10^3\,$erg$\,$cm$^{-2}$$\,$s$^{-1}$, and $\beta=1/2$ for $F_{\rm XUV}\geq 3\times10^3\,$erg$\,$cm$^{-2}$$\,$s$^{-1}$. Using the best-fit value of $\gamma \simeq 1/2$ found previously, we obtain  $\mathcal{P_{\rm AW}}\propto F_{\rm XUV}^{1/2}$ in the energy-limited regime and $\mathcal{P_{\rm AW}}\propto F_{\rm XUV}^{1/4}$ in the recombination-limited regime. The red solid line in the right panel of Figure \ref{fig:power_mdot_flux} shows this broken power-law fit, which gives a good match to the simulated data.

Last, we examine the contributions from thermal and kinetic energy fluxes to the total power transmitted towards the star. To isolate the power transmitted to the star, we consider the same analysis plane $\mathcal{A}$ chosen for the calculation of $\mathcal{P_{\rm AW}}$. The resulting kinetic energy flux is: 
\begin{equation}\label{eq:F_k}
    P_{\rm kin} = \int_{\mathcal{A}_+} \frac{1}{2} \rho u^2 \mathbf{u}\cdot \rm d\mathbf{\mathcal{A}_+} \quad  \mathcal{A}_+ = \{\mathbf{x}\in \mathcal{A}: \mathbf{u}(\mathbf{x})\cdot\mathbf{c^-_A} >0\}.
\end{equation}
Similarly, the thermal power is given by
\begin{equation}\label{eq:F_th}
    P_{\rm th} = \int_{\mathcal{A}_+} \frac{5}{2} \frac{\rho k_B T}{m_H} \mathbf{u}\cdot \rm d\mathbf{\mathcal{A}_+},
\end{equation}
where $T$ is the plasma temperature, $m_H$ is the hydrogen mass, and $k_B$ is the Boltzmann constant. 

Figure \ref{fig:power_mdot_flux} includes the thermal and kinetic contributions to the total power as orange and green circles, respectively. For $\dot{M}_d\lesssim 2\cdot10^{11}\,$g/s, the kinetic and thermal fluxes are either zero or negligible compared to the Poynting flux, in agreement with \citet{Saur13}. For higher mass-loss rates, the planetary wind becomes dynamically important, and the velocity vector points towards the star in the regions filled with planetary plasma. In those cases, the combined kinetic plus thermal power can reach about $10^{25}\,$erg/s.  

\section{Energy flux dependence on planetary magnetic field strength}\label{sec:4}
The power released by the star-planet interaction via Alfvén wings scales with the size of the planetary obstacle. More extended obstacles produce wider Alfvén wings that channel more Poynting flux towards the star. For magnetised planets, the obstacle comprises the planet itself and its magnetosphere. Therefore, to first order, the effective area of the star-planet magnetic interaction is proportional to the area of the magnetospheric cross-section \citep[e.g.][]{lanza09,vidotto23}. Since the magnetospheric size increases with the magnetic field strength of the planet, one expects that the power radiated to the star to be an increasing function of $B_p$. To quantify this aspect of star-planet magnetic interactions, we run several models with $B_p \in [2.5,5,10,25]\,$G while keeping the same stellar wind properties as in Section \ref{sec:3}. The relative alignment between the stellar and planetary fields also determines the effective area of the interaction \citep{Saur13,strugarek15,strugarek16,strugarek17}. Therefore, we consider three different angles between the stellar and planetary fields: $\Theta_M\in[0,45,90]^\circ$. 

\begin{figure*}
    \centering
    \includegraphics[width=0.999\textwidth]{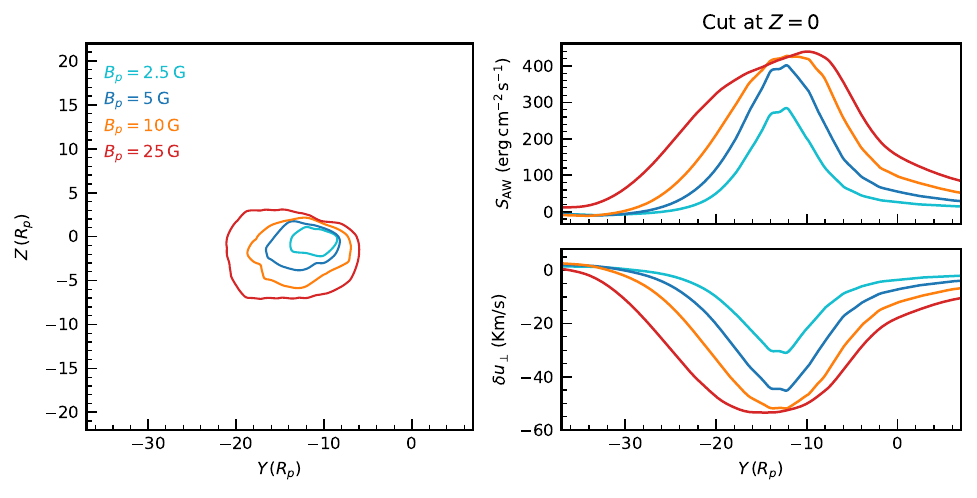}
    \caption{Similar to Figure \ref{fig:Poynting_plane}, for different planetary magnetic field strengths $B_p$ and $\Theta_M=90^\circ$. Left: boundary of the inner Alfvén wing, characterised by the cross-section of the flux tube connecting the star and the planet. Each contour is color-coded by magnetic field strength. The cross-section is taken at $X=-22\,R_p$ for the $B_p=2.5\,$G planet, and at $X=-24\,R_p$ for all of the other $B_p$. Right: properties of the Alfvén wing along the $Z=0$ line, in the same cross section as in the left panel of the Figure. The top panel shows the Poynting flux $S_{\rm AW}$, and the bottom panel represents the perturbation in velocity associated with the traveling Alfvén waves. Line colors in the right panels match the $B_p$ contour colors in the left panel.}
    \label{fig:Bp_poynting}
\end{figure*}

The left panel of Figure \ref{fig:Bp_poynting} shows the Alfvén wing cross-section for the different planetary magnetic field strengths considered. The dipolar axis is perpendicular to the (radial) stellar magnetic field in all cases ($\Theta_M=90^\circ$). The $YZ$ plane cut was taken at $X=-22,R_p$ for the $B_p=2.5\,\mathrm{G}$ model, and at $X=-24\,R_p$ for the remaining models. The Alfvén wing cross-section presents a roughly elliptical shape whose area increases with $B_p$. Similar to Figure \ref{fig:Poynting_plane}, we also plot the Poynting flux $S_{\rm AW}$ and Alfvénic velocity perturbation $\delta u_\perp$ along the $Z=0$ line in the top-right and bottom-right panels of Figure \ref{fig:Bp_poynting}, respectively. Both $S_{\rm AW}$ and $\delta u_\perp$ peak within the wing center and decline smoothly to their background values. As $B_p$ increases, the region with the strongest Poynting flux and velocity perturbation becomes more extended due to the larger extent of the wing. 

\begin{figure}
    \centering
    \includegraphics[width=0.49\textwidth]{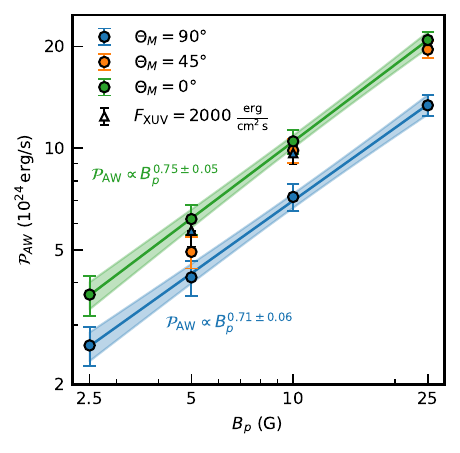}
    \caption{Alfvén wing power transmitted to the star as a function of planetary magnetic field strength. The circles represent the average power at several distances from the planet, along the $c_A^-$ wing, and the error-bars indicate the standard deviation of the mean. The solid lines are power law fits, and the shaded region around the lines represent the $1 \sigma$ region of the fit. The points and lines are colored by the relative magnetic field angle $\Theta_M$. Circles represent the models with XUV flux of 500 erg$\,$cm$^{-2}\,$s$^{-1}$, while triangles represent the 2000 erg$\,$cm$^{-2}\,$s$^{-1}$ cases.}
    \label{fig:Bp_power}
\end{figure}

To examine how the total power changes with $B_p$, we follow Equation \eqref{eq:power} and average $S_{\rm AW}$ over the Alfvén wing cross-section for each of our models. In Figure \ref{fig:Bp_power}, we show the power $\mathcal{P}_{\rm AW}$ as a function of $B_p$.  We integrate $S_{\rm AW}$ between $18$ 
and $22\,R_p$ for the $2.5\,$G models, and between $22$ and $26\,R_p$ for the other planetary magnetic field strengths. The circles and error-bars in the Figure indicate the mean power and its standard deviation, respectively. As expected, for each relative magnetic inclination $\Theta_M$, the power increases with $B_p$ due to the larger magnetospheric size. 

To understand the dependence of the Alfvén wing power on the magnetic field strength of the planet, we fitted a power law in the form $\mathcal{P}_{\rm AW}\propto B_p^{\alpha}$. We show our best fit in Figure \ref{fig:Bp_power} (solid lines), where the shaded area represents the $1\sigma$ confidence interval of the fit. We find power-law exponents in the range $\alpha \approx 0.7 - 0.75$, slightly higher than the theoretical expectation $\alpha=2/3$ derived from the scaling of power with the cross-sectional area of a dipolar magnetosphere. However, the difference is small: the theoretical exponent lies close to the lower edge of our $1\sigma$ interval, so our numerical results are broadly consistent with the theoretical trend. We verified that, in our simulations, the magnetospheric size follows the expected scaling $R_M \propto B_p^{1/3}$ (see Appendix \ref{appendix:magnetosphere}, where we also explain the origin of the $1/3$ exponent). The slightly steeper relation we find in our powers could also suggest that the effective area of the interaction extends beyond the magnetospheric cross-section. One idea is that the flux tube connecting the star and the planet also perturbs the stellar wind around it, and it therefore should be considered as part of the planetary obstacle \citep[see e.g.][]{chane12,strugarek16}. Another reason could be that planets with stronger magnetisation slow down the stellar wind more efficiently due to $\mathbf{j}\times\mathbf{B}$ forces ($\mathbf{j}$ is the current density). However, we remark that our scaling law with $B_p$ is in close agreement with the theoretical Alfvén wing model prediction of \citet{Saur13}.

We also find that, for a given planetary magnetic field strength, more aligned magnetic configurations with respect to the stellar wind field ($\Theta_M=0$, i.e., when the positive  planetary dipolar axis points towards the star) result in higher powers (see the different colors in Figure \ref{fig:Bp_power}).  
This behavior has been found in numerical simulations \citep{strugarek15,strugarek17} and predicted by analytical models \citep{Saur13}. It can be understood from the fact that more aligned magnetic configurations result in a larger connectivity between the planet and the stellar field, higher reconnection rates and more energy release. Here, we considered a perfectly aligned scenario ($\Theta_M=0^\circ$), a perpendicular case ($\Theta_M=90^\circ$), and an intermediate situation between the latter two ($\Theta_M=45^\circ$). In the next section, we will extend these results to additional inclinations.  

Lastly, in Figure \ref{fig:Bp_power} we show two additional models with $\Theta_M=90^\circ$ under an irradiation four times stronger than the baseline value of 500 erg$\,$cm$^{-2}\,$s$^{-1}$. These points are shown by blue triangles in the Figure. For this stellar flux, we observe that the power increases approximately by a factor of 2, so that it is comparable to a planet with the same magnetic field strength and a perfectly aligned field geometry with respect to the stellar wind. We note that, for stronger XUV fluxes, the power will increase even more (see Section \ref{sec:3}).  

\section{Energy flux dependence on magnetic field topology}\label{sec:5}
\begin{figure*}
    \centering
    \includegraphics[width=0.999\textwidth]{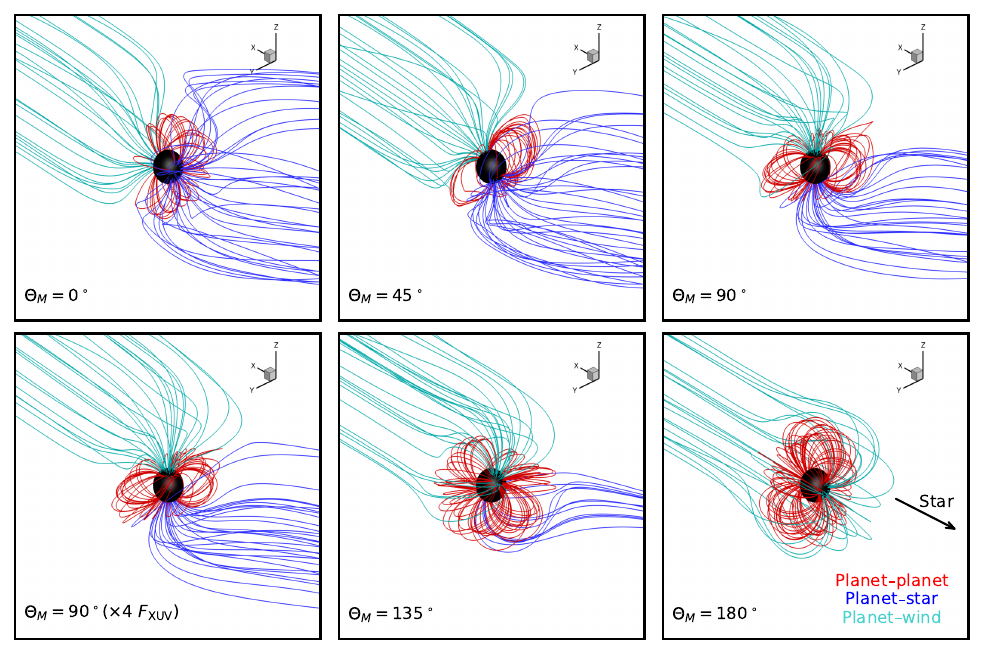}
    \caption{Magnetic field structure of a $B_p=5\,$G planet (central black sphere) under different magnetic inclinations with respect to the stellar wind magnetic field. The magnetic configurations range from completely aligned fields (magnetic inclination of $\Theta_M=0^\circ$) to completely anti-aligned fields ($\Theta_M=180^\circ$). The bottom-left panel displays a perpendicular magnetic configuration ($\Theta_M=90^\circ$) for a planet that receives 4 times the irradiation considered for all of the other models. 
    In all panels, the magnetic field lines of the planet are colored in blue if they connect back to the star (indicated by the black arrow, outside the computational grid), in cyan if they connect with the stellar wind away from the star, and in red if they remain closed.   }
    \label{fig:magnetic_topology}
\end{figure*}

To further investigate how the alignment between the stellar and planetary fields impacts the power released in the magnetic interaction, we consider a $5\,\rm G$ planet under different magnetic topologies ranging from a perfectly aligned configuration ($\Theta_M=0^\circ$) to a completely anti-aligned one ($\Theta_M=180^\circ$). We illustrate the different cases in Figure \ref{fig:magnetic_topology}, where we show a collection of field lines that originate from the planetary surface. The figure shows that the flux tube connecting the star and the planet, indicated by the blue field lines, is bigger for more aligned topologies. In the completely anti-aligned case ($\Theta_M=180^\circ$, bottom-right panel), the planetary field polarity is reversed, and all open planetary field lines are diverted to the night-side, where they merge with the stellar wind field (turquoise lines). The closed magnetosphere, indicated in red, has a similar size in all the cases shown in the Figure.

In the bottom-left panel of Figure \ref{fig:magnetic_topology}, we present a perpendicular configuration ($\Theta_M=90^\circ$) exposed to a stellar irradiation of $F_{\rm XUV} = 2000\,$erg$\,$cm$^{-2}\,$s$^{-1}$, 4 times the irradiation considered for all other models. Compared with the base-irradiation case (top-right panel of Figure \ref{fig:magnetic_topology}), the strongly irradiated model exhibits a broader flux tube connecting the star. This is because the outflow exerts an additional pressure from the inside of the flux tube, increasing the effective size of the interaction. As discussed in Section \ref{sec:3}, the outflow also perturbs the Alfvén wing from within, generating additional Poynting flux. These effects result in higher SPMI power.

\begin{figure}
    \centering
    \includegraphics[width=0.49\textwidth]{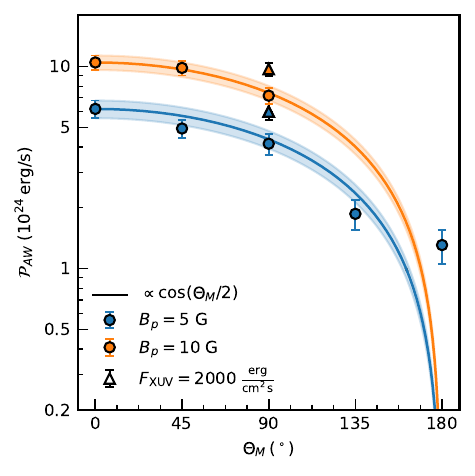}
    \caption{Alfvén wing power for different magnetic inclinations $\Theta_M$ and $F_{\rm XUV} = 500\,$erg$\,$cm$^{-2}\,$s$^{-1}$. The colors indicate the magnetic field strength of the planet. solid lines show the analytic dependence given by Equation \eqref{eq:power_thetaM_saur}. The shaded band around each line indicates the analytical trend for the $\pm 1\sigma$ values of $\mathcal{P_{\rm AW}}(\Theta_M=0^\circ)$. The triangles correspond to models with an incident irradiation of $F_{\rm XUV} = 2000\,$erg$\,$cm$^{-2}\,$s$^{-1}$  }
    \label{fig:theta_power}
\end{figure}

To compare how the power changes with the magnetic topology of the system in the 6 cases presented above, we show the Alfvén-wing power as a function of the magnetic inclination angle $\Theta_M$ in Figure \ref{fig:theta_power} (blue points). The Alfvén-wing power decreases with the magnetic inclination angle, and it is about 5 times higher in the perfectly aligned case compared to the completely anti-aligned scenario. This result is in agreement with a recent numerical study considering dipolar stellar fields \citep{Paul25}. According to the theoretical model of \cite{Saur13}, the effective radius of the interaction scales with the average width of the flux tube connecting the star and the planet, which is given by Equation \eqref{eq:Reff:saur}. Therefore, for a fixed magnetospheric standoff distance $R_m$, the SPMI power is expected to decrease with the magnetic inclination $\Theta_M$ as
\begin{equation}\label{eq:power_thetaM_saur}
    \mathcal{P}_{\rm AW}(\Theta_M) = \mathcal{P_{\rm AW}}(\Theta_M=0^\circ) \cos{(\Theta_M/2)}.
\end{equation}

We represent the theoretical dependence given by Equation \eqref{eq:power_thetaM_saur} by solid lines in Figure \ref{fig:theta_power}, colored by planetary magnetic field strength. Our results show that the \cite{Saur13} scaling is a good approximation for moderate magnetic inclinations ($\Theta_M \in [0^\circ,90^\circ]$), as seen also in the  $B_p=10\,$G cases displayed in the figure (orange circles). The analytical approximation is also close to the simulated value for the more anti-aligned case of $\Theta_M=135^\circ$, although it is not applicable to the completely anti-aligned case ($\Theta_M = 180^\circ$) since it would yield a null effective interaction size. In reality, even if no field lines connect back to the star, the magnetospheric obstacle would still perturb the stellar wind and the SPMI power would be non-zero. This is seen in the $\Theta_M=180^\circ$ model of Figure \ref{fig:theta_power} and other numerical studies \citep[e.g.][]{strugarek15,strugarek16,strugarek17}.

Lastly, the triangles in Figure \ref{fig:theta_power} show the power for two $\Theta_M=90^\circ$ models under an enhanced stellar irradiation of $F_{\rm XUV} = 2000\,$erg$\,$cm$^{-2}\,$s$^{-1}$. As seen in the previous Sections,
the powers found in the enhanced irradiation cases are higher than the scenarios with no-dayside escape conditions.  
 
\section{Discussion}\label{sec:6}
\subsection{Comparison with analytical estimates}\label{sec:analytical_comparison}
The total electromagnetic power transmitted along an Alfvén wing connecting the planet with the star is given by Equation \eqref{eq:power_aw_saur}. This analytical model from \citealt{Saur13} is a simplified expression derived under the assumption of small Mach numbers and a spatially homogeneous stellar wind at the position of the planet, although it remains a good approximation of the total Alfvén wing power for larger $M_A$ (an exact formulation valid for arbitrary $M_A$ is also given in their Equation~54). Here, we compare the powers computed directly from our simulations with the simplified expression of the Alfvén wing model given in Equation \eqref{eq:power_aw_saur}.

\begin{figure*}
    \centering    
    \includegraphics[width=0.999\textwidth]{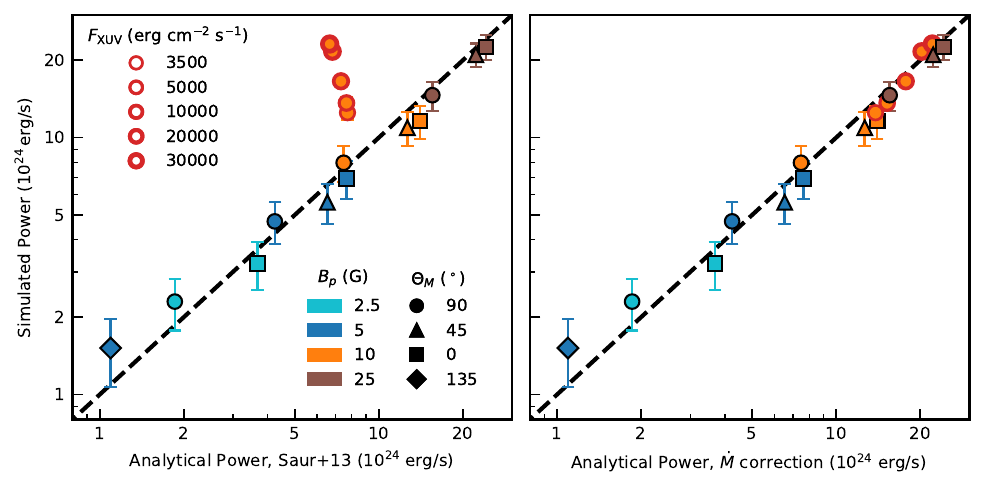}
    \caption{Alfvén wing powers from our simulations compared to those calculated using the analytical model given by Equation \eqref{eq:power_aw_saur} (left), and Equation \eqref{eq:p_aw,esc} (right). The dashed black line  indicates equal values. The points are colored by planetary magnetic field strength, symbols indicate different magnetic inclination. The models with enhanced stellar irradiation are indicated by red edges, the edge thickness increases with $F_{\rm XUV}$. Black edges indicate the fiducial $F_{\rm XUV}=500\,$erg$\,$cm$^{-2}\,$s$^{-1}$ case. 
    Left: analytical model without correction for atmospheric escape effects. Right: analytical model including a correction for the evaporation-driven enhancement of the open magnetic flux.}
    \label{fig:comparison_Bp}
\end{figure*}

This comparison is shown in the left panel of Figure \ref{fig:comparison_Bp}, where we calculated the analytical power using Equation \eqref{eq:power_aw_saur} with the unperturbed stellar wind values listed in Table \ref{tab:wind_parameters}. 
The effective size of the interaction was computed from Equation \eqref{eq:Reff:saur}, using the magnetospheric sizes given in Table \ref{tab:Rm_methods}. We calculated the interaction strength $\alpha$ using Equation \eqref{eq:alpha_saur}. In our simulations, the wind speed in our simulations drops from $\sim 150$ km/s to a few tens of km/s. This results in interaction strengths in the range $0.8-1$, where higher-$B_p$ models are better represented by $\alpha=1$, while models with weaker $B_p$ are closer to $\alpha=0.8$. We attribute this trend to Lorentz forces, which decrease the velocity of the stellar wind in the vicinity of the planet and scale with $B_p$. The interaction strength for all of our models is listed in Table \ref{tab:summary}.

In Figure \ref{fig:comparison_Bp}, the dashed black line marks the region where the analytical and numerical powers coincide. Most models show good agreement with the analytical prediction, consistent with previous MHD studies \citep[e.g.][]{strugarek15,fisher22}. However, the analytical \citet{Saur13} model does not include the effects of atmospheric escape, and it underestimates the power for the simulations that include enhanced stellar irradiation (points with thicker edges in Figure \ref{fig:comparison_Bp}).  
To account for the effects of atmospheric evaporation, we recompute the analytical power by including the extended Alfvén wing model given in Equation \eqref{eq:P_AW_Mdot}. Here, we adopted the best-fit power law exponent $\gamma=1/2$ derived in Section \ref{sec:3.2}, and calculated the scaling factor $\dot{M_0}$ using Equation \eqref{eq:M_lower}. The right panel of Figure \ref{fig:comparison_Bp} shows the comparison between the numerically derived powers and the  analytical model including the effects of atmosphere escape. When including the correction factor (Equation \eqref{eq:P_AW_Mdot}), the analytical prediction aligns more closely with the numerical results. This highlights the need to account for the effects of atmospheric escape when assessing SPMI energetics in strongly irradiated giant exoplanets.

\subsection{Non-ideal MHD effects: electrical resistivity}\label{sec:resistivity}
In the magnetised hot Jupiters considered in this work, the strength of the sub-Alfvénic interaction is mainly controlled by the coupling efficiency between the stellar wind and planetary fields, which is in turn governed by the strength and topology of both fields (see Sections \ref{sec:4} and \ref{sec:5}). The conductive properties of the planet and the stellar wind could also affect the coupling efficiency, potentially reducing the  power transferred towards the star. 

These effects are not included in our modelling approach and would require resistive MHD simulations to assess their impact in detail, which lies beyond the scope of this work. Nevertheless, their possible importance can be estimated by considering the interaction strength of a partially conducting ionosphere with Pedersen conductance $\Sigma_P$ \citep{neubauer1980,Saur13}:
\begin{equation}\label{eq:alpha_ionosphere}
    \alpha = \frac{\Sigma_P}{\Sigma_p + 2\Sigma_A}.
\end{equation}
Here, $\Sigma_A$ is the Alfvén conductance of the stellar wind, which sets the maximum current carried by the Alfvén wing. In SI units, it can be written as \citep{Saur13}
\begin{equation}\label{eq:Alfven_conductance}
    \Sigma_A = \frac{1}{\mu_0 u_A(1+M_A^2-2M_A \sin{\theta})^{1/2}},
\end{equation}
where $\mu_0$ is the vacuum magnetic permeability. To estimate the Pedersen conductance of our planet, we employ the power law scaling from \citet{nichols16}
\begin{equation}\label{eq:pedersen_conductance}
    \Sigma_P = 15.475 \Big(\frac{d}{1\, \rm au}\Big)^{-2.082}\Big(\frac{B_J}{B_p}\Big) \Big( \frac{L_{\rm XUV}}{L_{\rm XUV,\odot}} \Big)^{1/2} \quad \rm S,
\end{equation}
where $d$ is the orbital distance of the planet, $B_J$ is the Jovian magnetic field strength, and $L_{\rm XUV} / L_{\rm XUV,\odot}$ is the XUV luminosity of the star relative to the solar value. Equation \eqref{eq:pedersen_conductance} accounts for the conductivity generated by stellar irradiation, introducing an explicit dependence on both orbital distance and XUV luminosity. The power-law indices and the prefactor were chosen to reproduce Jovian values and conductance models for hot Jupiters.

For hot Jupiters with $d=0.01 - 0.1\,$ au and $B_p = 0.1 -10\,B_J$, and stellar XUV luminosities between $0.1-10\,L_{\rm XUV,\odot}$, Equation \eqref{eq:pedersen_conductance} predicts $\Sigma_P \sim 10^2 - 10^6\,$S. For typical stellar wind Alfvén speeds $u_A = 10^2 - 10^3\,$km/s, the Alfvén conductance is $\Sigma_A \sim 1-10\,$S. In this regime, $\Sigma_P \gg \Sigma_A$, implying a strong coupling and $\alpha$ close to unity for most hot Jupiter systems. Resistive effects may become important in other interaction regimes, particularly for satellites embedded within planetary magnetospheres \citep[e.g.][]{Saur13,blocker20}.  

\subsection{Application to an observed system: HD 189733}\label{sec:6.2}

Here, we present a practical application of the extension of the Alfvén wing model for evaporating planets presented in this work. We consider the HD 189733 star-planet system, consisting of a hot Jupiter orbiting a K0V star. This system 
exhibits both signatures of SPMI at the host star \citep{shkolnik08,cauley18,cauley19} and several atmospheric escape detections \citep{lecavalierdesetangs2010,bourrier13,dosSantos23}.

By analysing the excess absorption in the  Ca II K band of the star, \citet{cauley19} found a signal modulated close the orbital period of the planet corresponding to a power of about $5\times10^{26}\,$erg/s. By using the contemporaneous magnetic maps of the star obtained by \citet{Fares17}, \citet{strugarek22} modelled the stellar wind properties at the orbital location of the planet. They found that the Alfvén wing model did not produce a strong enough interaction to explain the observed signal, as it reaches powers of at most $10^{26}\,$erg/s.

To evaluate whether atmospheric escape from the hot Jupiter could enhance the Alfvén wing power, we first compute the lower mass-loss rate limit $\dot{M}_0$ from Equation \eqref{eq:M_lower}. Using the system parameters listed in Table \ref{tab:HD189733_params}, and adopting $x=0.1$ and $u_{\rm sw}/u_{\rm pw}=10$, we obtain $\dot{M}_0 \simeq 6.5\cdot 10^9\,$g/s. The Alfvén wing power will exceed the theoretical value of \citealt{Saur13} whenever the dayside mass-loss rate of the planet is larger than this limit ($\dot{M}_d > \dot{M}_0$). For HD 189733 b, estimates of the mass-loss rates based on the incident XUV flux and comparisons between hydrodynamic models and spectroscopic transit signatures give values in the range $\sim10^{10}-10^{12}\,$g/s \citep[e.g.][]{guo16,odert20,lampon21,dosSantos23}. 

\begin{table}
	\centering
	\caption{Properties of the HD 189733 system adopted in this work. The Poynting flux $S_{\rm AW}$ is the maximal value along the planetary orbit for the HJ07 stellar wind model in \citet{strugarek22}. The Mach number $M_A$, stellar wind background magnetic field $B_0$, and the stellar wind mass-loss rate $\dot{M}_\star$ are also taken from such model.}
	\label{tab:HD189733_params}
	\begin{tabular}{lrr} 
	\hline
	Parameter & Value & Reference \\
	\hline
  $M_\star$ ($M_\odot$) & 0.92 &  \citet{Bouchy05}  \\ 
  $R_\star$ ($R_\odot$) & 0.76 & \citet{Winn07}  \\
  $\dot{M}_{\star}$ ($M_\odot\,$yr$^{-1}$) & $2.6\times10^{-14}$ & \citet{strugarek22} \\
  $B_0$ (G) & 0.033 & \citet{strugarek22} \\
  $S_{\rm AW}$ (erg$\,$s$^{-1}$) & 830 & \citet{strugarek22} \\
  $M_A$ & 0.4 & \citet{strugarek22} \\
  $M_p$ ($M_{\rm J}$)       & 1.13 & \citet{Stassun17}  \\
  $R_p$ ($R_{\rm J}$)       & 1.13 & \citet{Stassun17}  \\
  $a_p$ (au)       & 0.03 & \citet{Stassun17} \\
  \hline
	\end{tabular}
\end{table}

Based in these studies, we consider total mass-loss rates between $\dot{M}_0=6.5\times 10^9\,$g/s and $2\times10^{12}\,$ g/s, and calculate the escape-enhanced Alfvén wing power from Equation \eqref{eq:P_AW_Mdot} for a set of planetary magnetic field strengths. We assumed that the dayside mass-loss rate is one half of the total, ($\dot{M}_d = \dot{M}/2$). We performed this calculation by adopting the stellar wind values reported by \citet{strugarek22} in their HJ07 model\footnote{Publicy available at \url{http://www.galactica-simulations.eu/db/STAR_PLANET_INT/HD189733_SPMI/}. We considered the $\alpha$$\beta$-NP case.}, given in Table \ref{tab:HD189733_params}, and using the best fit power-law index $\gamma=1/2$ derived in Section \ref{sec:3.2}. For each planetary magnetic field strength, we calculated the effective radius of the interaction using Equation \eqref{eq:Reff:saur} with $\Theta_M=0^\circ$ and $R_m = (B_p/B_0)^{1/3}R_p$. We note that this is equivalent to Equation \eqref{eq:Rm_analytical} when the total stellar wind pressure is dominated by the magnetic pressure (i.e. $P_{\rm total,sw}\simeq B^2_0 / 8\pi$). We also adopted a maximal interaction strength $\alpha=1$, as we expect that the planet has a highly conductive ionosphere (see Section \ref{sec:resistivity}).
Under these assumptions, the analytical Alfvén wing model given by Equation \eqref{eq:power_aw_saur} can be rewritten as \citep{strugarek22}
\begin{equation}\label{eq:power_saur_alt}
    \mathcal{P}_{\rm AW,th}=2\pi R^2_p S_{\rm AW} \cdot3M_a\left(\frac{B_0}{B_p}\right)^{-2/3},
\end{equation}
where $R_p$,$S_{\rm AW}$, $M_A$ and $B_0$ are listed in Table \ref{tab:HD189733_params}.

\begin{figure}
    \centering
    \includegraphics[width=0.999\linewidth]{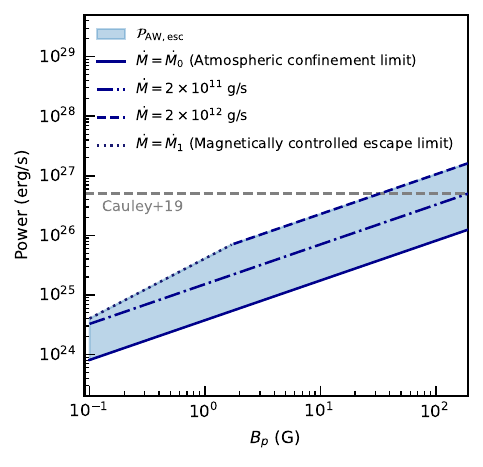}
    \caption{Maximum power generated by a star-planet magnetic interaction in the HD189733 system for different planetary magnetic field strengths $B_p$. The system parameters are given in Table \ref{tab:HD189733_params}. The solid blue line represents the power generated by the Alfvén wing model under negligible dayside escape conditions ($\mathcal{P_{\rm AW,th}}$), while the dotted line indicates the upper mass-loss rate limit $\dot{M}_1$ for magnetically controlled escape given by Equation \eqref{eq:M_1}. The shaded blue region represents the power released by the evaporation-enhanced Alfvén wing model ($\mathcal{P_{\rm AW,esc}}$) model, for total mass-loss rates between $\dot{M}_0$ and $2\times10^{12}\,$g/s. The dashed-dotted and dashed lines indicate the power for $\dot{M}=2\times10^{11}\,$g/s and $\dot{M}=2\times10^{12}\,$g/s, respectively.
    The observational signal detected by \citet{cauley19} is represented by the gray horizontal dashed line.}
    \label{fig:hd189733b}
\end{figure}

The resulting predictions are indicated with a shaded blue band in Figure \ref{fig:hd189733b}, where the solid blue line represents the theoretical \citet{Saur13} power $\mathcal{P}_{\rm AW,th}$. The dotted blue line marks the upper mass-loss limit $\dot{M}_1$ from Equation \eqref{eq:M_1}, above which 
atmospheric escape is no longer magnetically controlled, and the planet does not form a magnetosphere. In those cases, the power would correspond to the unmagnetised version of Equation \eqref{eq:power_saur_alt}, i.e. $B_0/B_p = 1$ and $R_m = R_p$.

Our results show a significant increase in the Alfvén wing power, which grows by a factor 4 for $\dot{M}=2\times10^{11}\,$g/s, and by a factor of 12 for $\dot{M}=2\times10^{12}\,$g/s. To assess whether such high mass-loss rates are plausible, we estimate the XUV luminosity $L_{\rm XUV}$ of HD 189733 from the observed X-ray luminosity $L_X$ of the host star. Adopting the quiescent value of $L_X=10^{28.18}\,$erg$\,$s$^{-1}$ from \citet{benjaffel13} and the scaling relation given in equation 3 of \citet{sanz-forcada11}, we obtain $L_{\rm XUV} = 1.2 \times 10^{29}\,$erg$\,$s$^{-1}$. This corresponds to $F_{\rm XUV} = 4.8 \times 10^4\,$ erg$\,$cm$^{-2}\,$s$^{-1}$ at the planetary orbit. In this case, Equation \eqref{eq:mdot_rl} predicts a recombination-limited escape rate of $\dot{M}=3.7 \times 10^{11}\,$g/s. After accounting for the fact that only about half of the escaping material flows toward the star, a planetary field strength of $B_p \simeq 120\,$G would be required to reproduce the Ca II power reported by \citet{cauley19}.

Although this value is substantially lower than the kilogauss-level fields required by the standard Alfvén-wing model, it is still larger than current theoretical estimates. For example, using the magnetic-field scaling laws of \citet{christensen09} and \citet{reiners10} combined with the stellar heat deposition scenario of \citet{yadav17}, \citet{cauley19} inferred a field strength of $B_p=53\pm17\,$G for HD 189733b. This estimate is likely an upper limit, as it assumes that all of the heating required to inflate the planet is deposited within the dynamo region. Therefore, a field strength of $B_p\simeq120\,$G appears unlikely, unless the efficiency of stellar heat deposition into the dynamo region is higher than what is usually assumed.

During flaring episodes, \citet{pillitteri22} reported stellar X-ray luminosities as high as $L_X=10^{29.3}\,$erg$\,$s$^{-1}$. While we are not aware of any flare detection during the observations analysed by \citet{cauley19}, their detection coincides with an epoch of high stellar magnetic activity \citep{Fares17}. It is plausible that, during that period, the star reached X-ray luminosities of order of $10^{29}\,$erg/s, comparable to the high-activity levels reported by \citet{pillitteri22}. 

A luminosity of $L_X=10^{29.3}\,$erg$\,$s$^{-1}$ corresponds to $F_{\rm XUV} = 4.7 \times 10^5\,$ erg$\,$cm$^{-2}\,$s$^{-1}$ and $\dot{M}=1.2 \times 10^{12}\,$g/s. In that case, a planet with $B_p \simeq 50\,$G would generate enough power to explain the observed signal, which is more compatible with magnetic field scaling laws. From these calculations, we conclude that the evaporation-enhanced Alfvén wing model might not explain the observed power very frequently, and may instead require the system to be observed during an active state. This interpretation appears consistent with the fact that \citet{cauley19} detected the signal in only 1 out of 6 epochs.

We caution that even higher Alfvén wing powers might be needed to explain the total energy budget, as the Ca II power is only a fraction of the total SPMI power \citep{paul25a}, and the Alfvén waves can be partially reflected in an inhomogeneous stellar wind. In this case, other SPMI mechanisms may contribute to produce the required energy. We discuss some possibilities in the next section. 

\subsection{Perspectives for observations of SPMI}
The results presented in this work show that the power released in star-planet interactions can be affected by the planet's atmospheric escape. This has interesting consequences for interpreting observations of star-planet interactions. First, as shown in Section \ref{sec:3.2}, the power released in the interaction increases with the mass-loss rate of the planet. Previous works have attempted to infer planetary magnetic field strengths by comparing analytical SPMI theories with the power estimated from planet-induced emission signals \citep{cauley19,ilin25}. In these studies, the standard Alfvén wing model predicts powers at least one to several orders of magnitude below those inferred from observations, unless very large planetary field strengths are assumed. 

In this study, we demostrated that photoevaporation of the planetary atmosphere can increase the effective area of the interaction, rising the Alfvén-wing power above the standard prediction. For the hot Jupiter HD189733 b, the Alfvén wing power could increase by an order of magnitude for strong stellar irradiation conditions (see Section \ref{sec:6.2}). However, if, as recent studies suggest, only about $10\,\%$ of the total SPMI power is converted into chromospheric emission \citep{paul25a} in the Ca II line, the required increase of power to explain these observations could be even higher.
   
Other mechanisms might therefore contribute to the observed signals. \cite{Saur13} found that, for most exoplanets, kinetic and thermal energy fluxes are typically directed away from the star or are negligible compared with the magnetic energy flux. However, if a stream of evaporating planetary material reaches the star,
these contributions might be important. This was the case in two of the models presented in this work (see Figure \ref{fig:power_mdot_flux}).
The accretion scenario might be possible in the HD 189733 system \citep{colombo24}, although detailed modeling including both the planet and the star is needed to assess the energetics of the interaction in this case.   
Alternatively, other mechanisms have been proposed that can produce the larger releases of energy needed to explain the observed powers. For instance, \cite{lanza13} showed that the release of energy stored in a stressed magnetic loop connecting the star and the planet can yield powers $10^2-10^3$ times larger than the Alfvén wing estimate. However, this scenario requires the planet to reside in a potential field environment (see \citealt{Saur2017,vidotto23}), a condition not met in the HD189733 stellar wind models of \cite{strugarek22}. Another possibility is that the planet acts as a trigger for the emergence of new stellar magnetic flux by increasing the magnetic helicity until a threshold value is reached \citep{lanza18}. 

In the \citet{lanza18} flux-emergence scenario, it is not possible to infer the magnetic field strength of the planet from SPMI signals, as the planet only acts as a trigger for the energy release. By contrast, if the Alfvén wing or interconnecting loop mechanisms are in action, it is possible, in principle, to estimate $B_p$ from the observed power, provided the stellar wind properties are well constrained. In practice, this is difficult because degeneracies remain. For example (see Figure \ref{fig:comparison_Bp}), a $10\,$G planet with $\Theta_M=90^\circ$, a $5\,$G planet with $\Theta_M=0^\circ$, and a more irradiated $5\,$G planet with $B_p=5\,$G and $\Theta_M=90^\circ$ can all produce comparable powers. 

One practical way to test whether atmospheric escape is enhancing the planet-induced signal is to measure planetary evaporation through spectroscopic transits (see the recent review by \citealt{vidotto25}). This method alone can be used to provide constraints on the planetary magnetic field strength, as magnetic fields can affect the transit profile \citep{Carolan21,schreyer24,presa24}. In particular, detecting a strongly red-shifted absorption component indicates material moving towards the star. This configuration would enhance the SPMI signal, and the corrections to the total power found in this work become relevant. In the future, to search for and interpret magnetic star–planet interactions, contemporaneous multi-wavelength observations will be essential: spectropolarimetry to constrain the stellar magnetic field, chromospheric emission monitoring to search for planet-induced activity, and spectroscopic transits to detect and characterise the atmospheric properties of the planet.

Finally, transient events such as coronal mass ejections and flares produce rapid changes in the local space weather conditions within very short timescales \citep[e.g.][]{Hazra22,hazra25,Elekes25}. These phenomena can alter the upstream plasma density, magnetic field and flow speed, while also driving stronger atmospheric escape from the planet. Consequently, we expect transient events to affect the magnetic connectivity of the star–planet interaction, and its associated Poynting flux. In the future, robust confirmation of SPMI will require  densely sampled observational campaigns that capture this short-timescale variability. 

\section{Conclusions}
In this work, we have studied how atmospheric escape influences the power released in star-planet magnetic interactions. To assess the energetics of the interaction, we performed 3D radiation-MHD simulations of a hot Jupiter planet orbiting in a sub-Alfvénic stellar wind, and integrate the Poynting flux channeled towards the star. We ran a grid of models varying the magnetic field strength of the planet, its obliquity, and the incident stellar XUV flux.

Our results show the formation of magnetic structures known as Alfvén wings, which carry Poynting flux in the form of Alfvén waves. Due to the radial geometry of our modelled stellar wind, one of the wings is connected to the star, while the other extends away from it. When stellar wind pressure largely confines the planetary outflow on the dayside, we find that atmospheric escape does not affect the power channeled towards the star. In this regime, our numerical results are in agreement with the powers predicted by the Alfvén wing model of \citet{Saur13}.

Conversely, if some planetary material flows through the dayside, the magnetic flux in the Alfvén wing increases, and the power transmitted to the star exceeds the analytical prediction of \citet{Saur13}. By comparing the semi-analytical formalism of Section \ref{sec:3.1} with our numerical simulations, we found a correction to the Alfvén wing power model which takes into account atmospheric escape. This correction is given by
\begin{equation}\label{eq:p_aw,esc}
    \mathcal{P_{\rm AW, esc}=\mathcal{P}_{\rm AW,th}}\left(\frac{\dot{M}_d}{\dot{M_0}}\right)^{1/2} \;\; \text{for $\dot{M_d} \in (\dot{M_0}, \dot{M}_1)$},
\end{equation}
where $\mathcal{P}_{\rm AW,th}$ is the theoretical Alfvén wing power given by Equation \eqref{eq:power_aw_saur}, $\dot{M}_d$ is the dayside mass-loss rate of the planet, $\dot{M}_0$ is the scale factor given by Equation \eqref{eq:M_lower}, and $\dot{M}_1$ is the upper mass-loss rate limit for magnetically controlled escape given by Equation \eqref{eq:M_1}. The following steps can be taken to estimate the Alfvén wing power enhanced by atmospheric escape:
\begin{itemize}[leftmargin=*, labelsep=0.5em]
    \item[1] Estimate the incident XUV flux received by the exoplanet \citep[e.g. using scaling relations from X-ray emission, see][]{sanz-forcada11}, and compute the atmospheric escape properties of the exoplanet. The global mass-loss rate of the planet $\dot{M}$ can be estimated using the analytical Equations \eqref{eq:mdot_el}-\eqref{eq:mdot_rl}, or calculated through more detailed (magneto)hydrodynamic models. Assuming a spherically symmetric outflow, the dayside mass-loss rate of the planet is then $\dot{M}_d \approx \dot{M}/2$. The planetary outflow speed $u_{\rm pw}$ can also be extracted from hydrodynamic models. If no model is available, it may be approximated by the sound speed, with typical temperatures of $5-10\times10^3\,$K.
    \\  
    \item[2] Compute $\dot{M}_0$ through Equation \eqref{eq:M_lower} to assess if the planetary outflow is confined by the stellar wind on the dayside. This requires the stellar wind velocity $u_{\rm sw}$ and mass-loss rate $\dot{M}_\star$, which need to be derived from stellar wind models. 
    For an approximate description, simplified prescriptions such as the \citealt{parker58} or \citealt{WeberDavies67} wind models can be employed. Alternatively, more detailed polytropic \citep[e.g.][]{reville16} or Alfvén wave driven \citep[e.g.][]{bellotti24} winds can be applied.
    \\
    \item[3] If $\dot{M_d}\lesssim \dot{M_0}$, the total power is well approximated by the \citet{Saur13} Alfvén wing model, given by Equation \eqref{eq:power_aw_saur}. 
    \\
    \item[4] If $\dot{M_d} > \dot{M_0}$, the planetary outflow enhances the power transmitted to the star. Equation \eqref{eq:p_aw,esc} can be used to estimate the power released by the star-planet interaction, as long as $\dot{M}_d < \dot{M}_1$.
    For escape rates above this threshold, the planetary outflow radially opens the planetary field lines, the effective interaction size reduces to the planetary radius $R_p$, and the power decreases to the unmagnetised limit. For a given planetary magnetic field strength $B_p$, the upper mass-loss rate limit can be found using Equation \eqref{eq:M_1}.
\end{itemize}

Applying this method to the HD189733 system, we show that the magnetic energy flux can be between 4 and 12 times higher than the standard predictions for dayside mass-loss rates between $10^{11}$ and $10^{12}\,$g/s, respectively. This  implies that the planetary magnetic field strength required to reproduce the \citet{cauley19} observations would be substantially lower than that predicted by the standard Alfvén-wing model, which would require kilogauss-level fields.
If evaporation is strong, a stream of planetary material can also be channeled toward the star along magnetic field lines. As seen in Section \ref{sec:3.2}, kinetic and thermal fluxes can further raise the total energy budget. These results suggest that atmospheric evaporation can substantially modify SPMI energetics, and should be taken into account when interpreting and predicting SPMI signals from photoevaporating exoplanets.

\section*{Acknowledgements}
We thank the referee for their constructive comments about our work.
This publication is part of the project "Blowing in the wind: exoplanetary atmospheres being carried away by stellar winds" (with project number VI.C.232.041 of the research programme "NWO Talent Programme VICI 2023"), which is financed by the Dutch Research Council (NWO). AAV acknowledges funding from the European Research Council (ERC) under the European Union's Horizon 2020 research and innovation programme (grant agreement No 817540, ASTROFLOW). This work used the BATS-R-US tools developed at the University of Michigan Center for Space Environment Modeling and made available through the NASA Community Coordinated Modeling Center. This work used the Dutch national e-infrastructure with the support of the SURF Cooperative using grant nos. EINF-7488 and EINF-13632. AP acknowledges D. Evensberget and A. Falorca for discussions on magnetohydrodynamic waves, and M. Cavieres for constructive comments on  outflow geometries. 

\section*{Data Availability}
The data described in this article will be shared on reasonable request to the corresponding author.



\bibliographystyle{mnras}
\bibliography{bibliography} 




\appendix

\section{Model parameters and results}\label{appendix:summary_table}
In this Appendix, we present the main parameters and numerical results for all of our models. These are summarised in Table \ref{tab:summary}.
\begin{table}
\centering
\caption{Main parameters and numerical results of the simulations. $B_p$ denotes the dipolar surface magnetic field strength at the poles of the planet, $\Theta_M$ is the angle between the radial stellar wind field and the planetary dipolar axis, and $F_{\rm XUV}$ is the XUV flux received by the planet. $R_m$ represents the size of the planetary magnetosphere. $\langle \mathcal{P}_{\rm AW}\rangle$ and $\Delta\mathcal{P}_{\rm AW}$ correspond to the mean SPMI power along the Alfvén wing, and its dispersion, respectively. $\alpha$ is the strength of the sub-Alfvénic interaction.}

\label{tab:summary}
\begin{tabular}{ccccccc}
\hline
$B_p$ & $\Theta_M$ & $F_{\rm XUV}$ & $R_m$ & $\langle \mathcal{P}_{\rm AW}\rangle$ & $\Delta\mathcal{P}_{\rm AW}$ & $\alpha$\\
(G) & ($^\circ$) & (erg$\,$cm$^{-2}\,$s$^{-1}$) & ($R_p$) & ($10^{24}\,$erg/s) & ($10^{24}\,$erg/s) & - \\
\hline
2.5 & 0 & 500 & 2.7 & 3.2 & 0.7 & 0.8 \\
2.5 & 90 & 500 & 2.7 & 2.3 & 0.5 & 0.7 \\
5.0 & 0 & 500 & 3.5 & 6.9 & 1.2 & 0.9 \\
5.0 & 45 & 500 & 3.5 & 5.6 & 1.0 & 0.9 \\
5.0 & 90 & 500 & 3.5 & 4.7 & 0.9 & 0.8 \\
5.0 & 90 & 2000 & 3.5 & 6.0 & 0.5 & 0.9 \\
5.0 & 135 & 500 & 3.5 & 1.5 & 0.4 & 0.6 \\
5.0 & 180 & 500 & 3.5 & 1.0 & 0.4 & 0.5 \\
10.0 & 0 & 500 & 4.6 & 11.6 & 1.7 & 1.0 \\
10.0 & 45 & 500 & 4.6 & 10.9 & 1.6 & 0.9 \\
10.0 & 90 & 250 & 4.3 & 7.2 & 0.7 & 0.8 \\
10.0 & 90 & 500 & 4.3 & 8.0 & 1.3 & 0.9 \\
10.0 & 90 & 750 & 4.3 & 6.6 & 0.7 & 0.8 \\
10.0 & 90 & 1000 & 4.3 & 7.3 & 0.7 & 0.9 \\
10.0 & 90 & 1250 & 4.3 & 7.8 & 0.7 & 0.9 \\
10.0 & 90 & 1500 & 4.3 & 8.3 & 0.7 & 0.9 \\
10.0 & 90 & 1750 & 4.2 & 9.3 & 0.7 & 1.0 \\
10.0 & 90 & 2000 & 4.2 & 9.9 & 0.6 & 1.0 \\
10.0 & 90 & 2250 & 4.1 & 11.1 & 0.8 & 1.0 \\
10.0 & 90 & 3500 & 4.0 & 12.5 & 0.7 & 1.0 \\
10.0 & 90 & 5000 & 4.0 & 13.6 & 0.7 & 1.0 \\
10.0 & 90 & 10000 & 4.0 & 16.6 & 0.6 & 0.9 \\
10.0 & 90 & 20000 & 4.0 & 21.6 & 0.6 & 0.9 \\
10.0 & 90 & 30000 & 4.0 & 23.1 & 0.1 & 0.9 \\
25.0 & 0 & 500 & 5.9 & 22.5 & 2.5 & 1.0 \\
25.0 & 45 & 500 & 5.9 & 21.0 & 2.2 & 1.0 \\
25.0 & 90 & 500 & 5.9 & 14.6 & 1.9 & 0.9 \\
\hline
\end{tabular}
\end{table}

\section{Alfvén-wing Power as a function of distance from the planet}\label{appendix:power_distance}
\begin{figure}    
    \centering
    \includegraphics{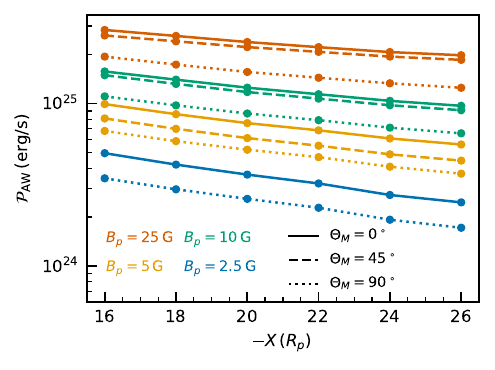}
    \caption{Alfvén wing power integrated at cross-sections placed at different distances $X(R_p)$ from the planet. Colors indicate different planetary magnetic field strengths, and each linestyle represents a different dipolar obliquity.}
    \label{fig:power_distance}
\end{figure}

Here we show the Alfvén wing power computed at several distances from the planet. As seen in Figure \ref{fig:power_distance}, there are small losses of power as the integration plane moves away from the planet that are likely due to numerical dissipation. The power decrease with distance is more severe for models with lower $B_p$, where the Alfvén wing is  narrower and harder to resolve. For higher magnetic field strengths the power is more stable with distance.  

\section{Magnetospheric size}\label{appendix:magnetosphere}
\begin{figure}
    \centering
    \includegraphics{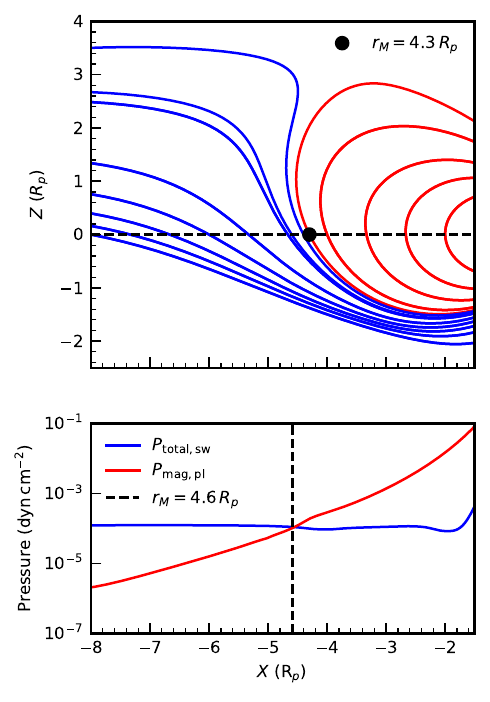}
    \caption{Magnetospheric size $R_m$ of a $10\,$G planet using two different approaches. (Top) Tracing the last closed-field line of the planet (red lines) along the star-planet line. (Bottom) Balancing the total pressure of the stellar wind (blue line) against the magnetic pressure of the planet (red line). }
    \label{fig:Rm_calculation}
\end{figure}

We calculate the size of the magnetosphere $R_m$ by tracing the last closed field line of the planet in a line perpendicular to the dipole axis, in the upstream direction. We illustrate this method in the top panel of Figure \ref{fig:Rm_calculation}, for a $B_p = 10\,$G, $\Theta_M = 90^\circ$ planet.  
Alternatively, $R_m$ can be estimated by balancing the total stellar-wind pressure $P_{\rm total,sw}$ (ram + thermal + magnetic) against the magnetic pressure of the planet $P_{\rm mag,pl}$, measured along the same direction. This pressure equilibrium is shown in the lower panel of Figure \ref{fig:Rm_calculation}. Table \ref{tab:Rm_methods} lists the magnetospheric distances obtained with both methods for each planetary magnetic field strength. Both methods yield very similar values, and using the $R_m$ found from pressure balance instead of the line-tracing values changes the analytical Alfvén wing power reported in Section \ref{sec:analytical_comparison} by less than $10\, \%$.
\begin{table}
\centering
\caption{Magnetospheric size $R_m$ obtained with two methods for different planetary field strengths and $\Theta_M=90^\circ$. The results are almost identical for the other magnetic obliquities. }
\label{tab:Rm_methods}
\begin{tabular}{lcc}
\hline
$B_p\,$(G) & $R_m/R_p$ (pressure balance) & $R_m/R_p$ (line tracing) \\
\hline
2.5  & 2.8 & 2.7 \\
5    & 3.7 & 3.5 \\
10   & 4.6 & 4.3 \\
25   & 6.1 & 5.9 \\
\hline
\end{tabular}
\end{table}

\begin{figure}
    \centering
    \includegraphics{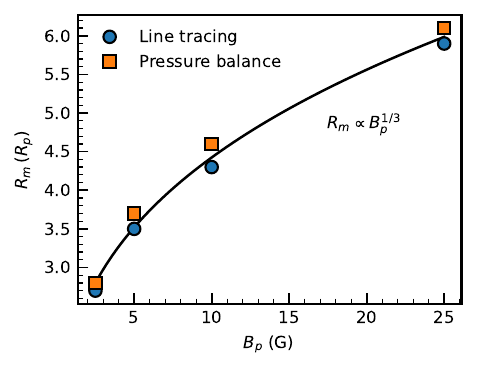}
    \caption{Planetary magnetospheric sizes listed in Table \ref{tab:Rm_methods} as a function of the magnetic field strength at the pole of the planet $B_p$. The solid black line represents the expected trend for a dipolar field.}
    \label{fig:RmvsRp}
\end{figure}

For a planetary dipolar field, the magnetic pressure acting against the stellar wind decreases with distance as $P_{\rm mag,pl}(r) = B^2_p(R_p / r)^{6}/(8\pi)$. Equating $P_{\rm mag,pl}(R_m)$ with the stellar wind pressure $P_{\rm total,sw}$ gives the magnetospheric size:
\begin{equation}\label{eq:Rm_analytical}
    R_m / R_p = (8\pi\,P_{\rm total,sw})^{-1/6}B^{1/3}_p.
\end{equation}
Equation \eqref{eq:Rm_analytical} shows that, for a dipolar planetary field, $R_m$ should scale with the planetary field strength as $B_p^{1/3}$. Figure \ref{fig:RmvsRp} plots the magnetospheric radii from Table \ref{tab:Rm_methods} against 
$B_p$. Our results closely follow the expected scaling $R_m \propto B_p^{1/3}$.  

\bsp	
\label{lastpage}
\end{document}